\documentclass[%
 reprint,
 amsmath,amssymb,
 aps,
prb,
]{revtex4-2}
\usepackage{chemformula} 
\usepackage[T1]{fontenc} 
\usepackage{bm}
\usepackage{amsmath}
\usepackage{amssymb}
\usepackage{amsthm}
\usepackage{amsfonts}
\usepackage{enumerate}
\usepackage{latexsym}
\usepackage{ifpdf}
\usepackage{graphicx}
\usepackage{makeidx}
\usepackage{color}
\usepackage{times}
\usepackage{chemformula} 
\usepackage[T1]{fontenc} 

\begin{document} 

\title{Orientation-dependent stabilization of MgCr$_2$O$_4$ spinel thin films }

\author{Fangdi Wen}
\email{fangdi.wen@rutgers.edu}
\affiliation{Department of Physics and Astronomy, Rutgers University, Piscataway, New Jersey 08854, USA}

\author{Xiaoran Liu} 
\affiliation{Department of Physics and Astronomy, Rutgers University, Piscataway, New Jersey 08854, USA}

\author{Mikhail Kareev} 
\affiliation{Department of Physics and Astronomy, Rutgers University, Piscataway, New Jersey 08854, USA}

\author{Padraic Shafer}
\affiliation{Advanced Light Source, Lawrence Berkeley National Laboratory, Berkeley, California 94720, USA}

\author{Tsung-Chi Wu} 
\affiliation{Department of Physics and Astronomy, Rutgers University, Piscataway, New Jersey 08854, USA}

\author{Michael Terilli} 
\affiliation{Department of Physics and Astronomy, Rutgers University, Piscataway, New Jersey 08854, USA}

\author{Elke Arenholz}
\affiliation{Advanced Light Source, Lawrence Berkeley National Laboratory, Berkeley, California 94720, USA, Now at Cornell High Energy Synchrotron Source, Cornell University, Ithaca, NY 14853}

\author{Jak Chakhalian}
\affiliation{Department of Physics and Astronomy, Rutgers University, Piscataway, New Jersey 08854, USA}

\date{\today}

\begin{abstract}
AB$_2$O$_4$ normal spinels with a magnetic B site can host a variety of magnetic and orbital frustrations leading to spin-liquid phases and field-induced phase transitions. 
Here we report the first epitaxial growth of (111)-oriented MgCr$_2$O$_4$ thin films. 
By characterizing the structural and electronic properties of films grown along (001) and (111) directions, the influence of growth orientation has been studied. 
Despite distinctly different growth modes observed during deposition, the comprehensive characterization reveals no measurable disorder in the cation distribution nor multivalency issue for Cr ions in either orientation. 
Contrary to a naive expectation, the (111) stabilized films exhibit a smoother surface and a higher degree of crystallinity than (001)-oriented films. The preference in growth orientation is explained within
the framework of heteroepitaxial stabilization in connection to a significantly lower (111) surface energy. 
These findings open broad opportunities in the fabrication of 2D kagome-triangular heterostructures with emergent magnetic behavior inaccessible in bulk crystals.
\end{abstract}

\maketitle

\section{Introduction}
Geometrically frustrated magnets have received considerable attention, and great efforts have been put forward to identify and characterize frustration-induced phenomena \cite{spin2010LB,experimental2019JW,quantum2017YZ,Herbertsmithite2016MN}. 
In two dimensions (2D), many antiferromagnetic materials with a triangular lattice motif can harbor frustrated interactions, making them excellent candidates for exotic behavior including quantized magnetization plateaus\cite{Nearly2020SZ}, charge frustration in mixed-valence spinels\cite{Orbital2016RK,Charge2015AU}, order by disorder\cite{Defect2019AA}, valence-bond ordering\cite{Mott2007YS,Frustration2006MT}, 'molecule-like spin clusters\cite{Competing2005CS,Classification2012MB}, and potentially spin-liquid states. \cite{Exotic2019VOG,Quantum2016LS,Herbertsmithite2016MN,Physics2018MH,Concept2019HT,Quantum2020CB} 
Despite a plethora of theoretical predictions, only a few highly-frustrated candidates have been investigated in detail so far.\cite{Herbertsmithite2016MN,Quantum2016LS,quantum2017YZ,Quantum2020CB} 
For instance, in three dimensions (3D),  corner-sharing tetrahedral pyrochlores A$_2$B$_2$O$_7$ with spins coupled either ferromagnetically or antiferromagnetically have been proposed to host numerous interesting phenomena stemming from macroscopic ground state degeneracy. \cite{magnetic2010JG,Frustrated2019JR,Spin2001SB,spin2010LB,Artificial2013CN,Spin2019CC} In close analogy to pyrochlores, spinels (AB$_2$O$_4$) represent another materials family that exhibits a corner-sharing tetrahedron network, with the B site ions forming a pyrochlore sublattice and A site ions organized into moderately frustrated diamond sublattice. \cite{Frustrated2010SL,Antiferromagnetic1952YY,Magnetic2017JL,Deviation2019DR} 
Based on this observation, it is unsurprising that spinels also demonstrate many unusual low-temperature magnetic phenomena linked to the massive ground state degeneracy such as spin glass, spin ice, and spin liquid states. \cite{magnetic2010JG,Geometric2005NT,Melting2007TS,Spin2004VH,Frustrated2018JC,Emergent2019XL,Quantum2020XL}

In this work, we specifically focus on MgCr$_2$O$_4$ (MCO). MCO is a normal spinel that crystallizes into the cubic {\it{Fd$\bar{3}$m}} space group with a = 8.33\AA.\cite{Sensitivity2011SD,Low2008LO} 
In this spinel, the Cr$^{3+}$ (3d$^3$) ions are octahedrally coordinated by oxygen ions resulting in S = 3/2 high-spin state under the cubic crystal field. 
Due to the direct overlap between t$_{2g}$ orbitals, the antiferromagnetic exchange is dominant between nearest-neighbor Cr$^{3+}$ ions leading to the high Curie-Weiss temperature $\Theta_{CW}=-400$K.\cite{Low2008LO,effect2003DM}
On the other hand, the corner-sharing tetrahedra sublattice of Cr$^{3+}$ ions remain paramagnetic down to the transition temperature $T_N$=12.5K, yielding a moderate value of the frustration parameter f=$|\theta_{CW}|/T_N$ $\sim$ 32.
At $T_N$, the system undergoes a magneto-structural phase transition driven by the strong spin-lattice coupling. \cite{MT2002Muon,Spinel007HS,crystal2013MK,Low2008LO,Sensitivity2011SD} 
Several neutron scattering experiments have elucidated the underlying frustrated behavior leading to the discovery of a highly-frustrated spin texture and `proximate' spin liquid due to the short-range exchange interaction. \cite{Molecular2008KT,Emergence2013KT,Canted2014AM,magnetic2019SB} 

\begin{figure*}[t]
\vspace{-0pt}
\includegraphics[width=0.7\textwidth]{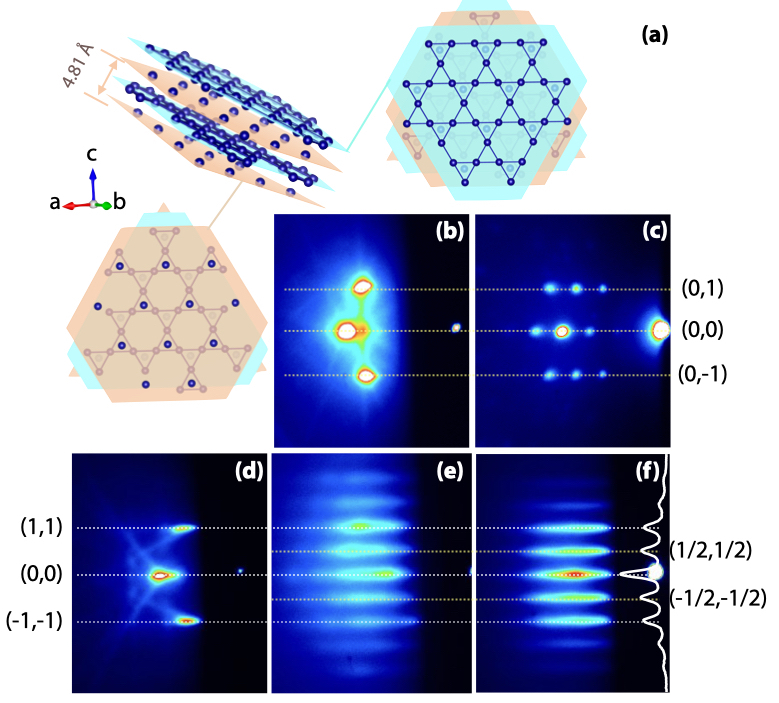}
\caption{\textbf{a.} Schematic  showing  Cr atoms in MCO when viewed along  (111) planes. \textbf{b.} RHEED pattern from the   (001) surface of MAO  before growth. \textbf{c.} RHEED pattern of MCO(001) grown on MAO(001). \textbf{d.} RHEED image of the AlO (0001)  substrate. \textbf{e.} RHEED pattern of MCO(111) grown on MAO(111). \textbf{f.} RHEED pattern of MCO(111) grown on AlO(0001). The inset curve is a vertical line-profile. All samples are grown at 550 $^\circ$C and  2 mTorr of pure oxygen. }
\label{RHEED}
\end{figure*}

In parallel  to  those developments, in pursuit of enhanced and tunable exchange interaction accompanied by strongly elevated magnetic frustration, a geometrical lattice engineering approach based on the synthesis of thin films of spinels along (111) becomes an increasingly popular direction. \cite{Frustrated2018JC,Emergent2019XL,Quantum2020XL,Initial2004UL,Strongly2020JC,Geometrical2016XL} 
To illustrate this approach, we note that along the (111) direction, Cr sublattice consists of alternating kagome and triangular atomic planes formed by magnetically active Cr$^{3+}$  ions (see Fig.\ref{RHEED}\textbf{a}). 
Thus the expectation is that by growing MCO(111) thin films, one can lower the dimensionality and shift the energy balance to activate stronger exchange interactions to potentially reach new magnetic states with frustrated behavior unattainable in bulk. 
Despite the intriguing proposals linked to the (111) geometry, to our best knowledge, the growth  of MCO thin films has not been demonstrated yet.
It is also interesting to note that in bulk a wide variety of methods have been applied to synthesize MCO crystals, including ceramic synthesis, co-precipitation, sol-gel method, combustion, hydrothermal method, zone melting, and co-crystallization. \cite{Structural1971NWG,magnesiochromate2020NFK} 
However, to date, the size of MCO crystals is still limited, which hinders the application of powerful experimental probes. \cite{Catalytic2014JH,Synthesis2016EJ,optical2013SK,Thermal2014SK,synthesis2010LM,bindi2014x}

In this work, we describe the growth of MCO(111) and MCO(001) thin films created by way of epitaxial stabilization using pulsed-laser deposition (PLD) with \textit{in-situ} reflection high-energy electron diffraction (RHEED). 
All the samples are characterized by X-ray reflectivity (XRR) and X-ray diffraction (XRD) using the Empyrean platform by Malvern Panalytical. 
X-ray photoelectron spectroscopy(XPS) was performed in a K-alpha X-ray photoelectron spectrometer (Thermo Fisher Scientific). 
Synchrotron-based X-ray absorption spectroscopy (XAS) was carried out on beamline 4.0.2 at the Advanced Light Source, Lawrence Berkeley National Laboratory. 
Based on the thorough analysis and contrary to the initial expectation, the results demonstrate that MCO growth on  (111)  surface is preferentially stabilized over the conventional (001) growth. 
This result was attributed to the significant difference in the orientation-dependent Gibbs energy during the initial phase of nucleation and growth. 
The successfully stabilized high-quality MCO(111) thin films expand the boundary of the materials phase space beyond bulk and potentially enable new magnetic states with frustrated behavior. 

\section{Experimental section}

A series of MCO films have been prepared on (100) and (111) oriented surfaces.
All samples were grown by pulsed laser deposition with the same laser power of $\sim$2 J/cm$^2$, pulse-rate of 4 Hz, and monitored by  {\it{in-situ}} RHEED. 
During the deposition, the best growth condition occurs for the substrate temperature of 550 $^\circ$C and 2 mTorr oxygen background pressure. 
After the growth, all films were annealed at the growth condition for 10 mins, and then cooled down to room temperature at 15 $^\circ$C/min under the same oxygen pressure. 
Particularly, to create (111)-oriented films, we focused on two different substrates. One is MgAl$_2$O$_4$ spinel (MAO), which has the same structural morphology and a lattice mismatch $\epsilon =( {a_{\textrm{MAO}}-a_{\textrm{MCO}}})/{a_{\textrm{MCO}}}$ $\sim -2.9$\% (compressive strain). 
As for the second choice, we selected $\alpha$-Al$_2$O$_3$(AlO), whose oxygen sublattice is in close registry with  MCO oxygen sublattice. 
Here we note, the in-plane averaged O-O distance of AlO is 2.75\AA~ while the O-O distance on the (111) surface of MCO is 2.95 \AA~, resulting in the \textit{continuity of anion sublattice} across the film/substrate interface albeit with a  larger compressive strain of $\sim$ - 6.8\%. 

In order to grow MCO (001), we used MAO (001) as a substrate. 
As shown in Fig. \ref{RHEED}\textbf{b} and 1\textbf{c}, after the deposition of MCO, sharp streaks in the RHEED pattern were observed with the vertical spacing matching the substrate (0,1) and (0,-1) reflections. 
This observation is typical for the films crystallized in the Stranski-Krastanov growth mode. 
However, after switching to the (111) surface of an otherwise identical MAO substrate, sharp RHEED peaks rapidly turn in to diffuse streaks along the (1,-1) axis, as illustrated in Fig. \ref{RHEED}\textbf{e}. 
This observation implies that after changing the substrate orientation from (111) to (001)\ , the growth mode is switched to the Frank-van-der-Merve (3D island) mode.
Surprisingly, despite the more substantial disparity in the crystal structure and larger strain, the AlO substrate serves as a better template for growing MCO(111). 
Figure \ref{RHEED}\textbf{d} and 1\textbf{f} compares the RHEED patterns of AlO (0001) substrate with  the MCO film. 
As clearly seen, the RHEED pattern of the substrate peaks (-1,-1) and (1,1) matches well with the  (-1,-1) and (1,1) RHEED reflections of the film. 
Since the in-plane lattice constant of MCO is twice of that of AlO, an extra pair of strong reflections appears almost immediately after the initial deposition [marked by ($\pm$1/2,$\pm$1/2) in Fig \ref{RHEED}\textbf{c}]. 
A direct comparison to Fig.  1\textbf{e} reveals much smoother streaked features with an overall lower background observed after deposition on the AlO substrate, confirming the Frank-van-der-Merve growth mode. 
These findings signify the critical importance of matching the anion network over the magnitude of strain. 
One can speculate that a possible reason for the observed island growth on MAO (111) substrate is that the polar MAO (111) surface  is compensated for the polar discontinuity with ordered oxygen vacancies leading to the reconstructed $(6\sqrt3\times6\sqrt3)$ areas. 
As a result, the charge compensated surface will consist of a juxtaposition of the original oxygen-terminated MAO (111) domains disrupted by oxygen-deficient areas that impede the MCO crystallization.
\cite{Noncontact2012BJN,structure2007ND} 

\begin{figure*}[t]
\vspace{-0pt}
\includegraphics[width=0.8\textwidth]{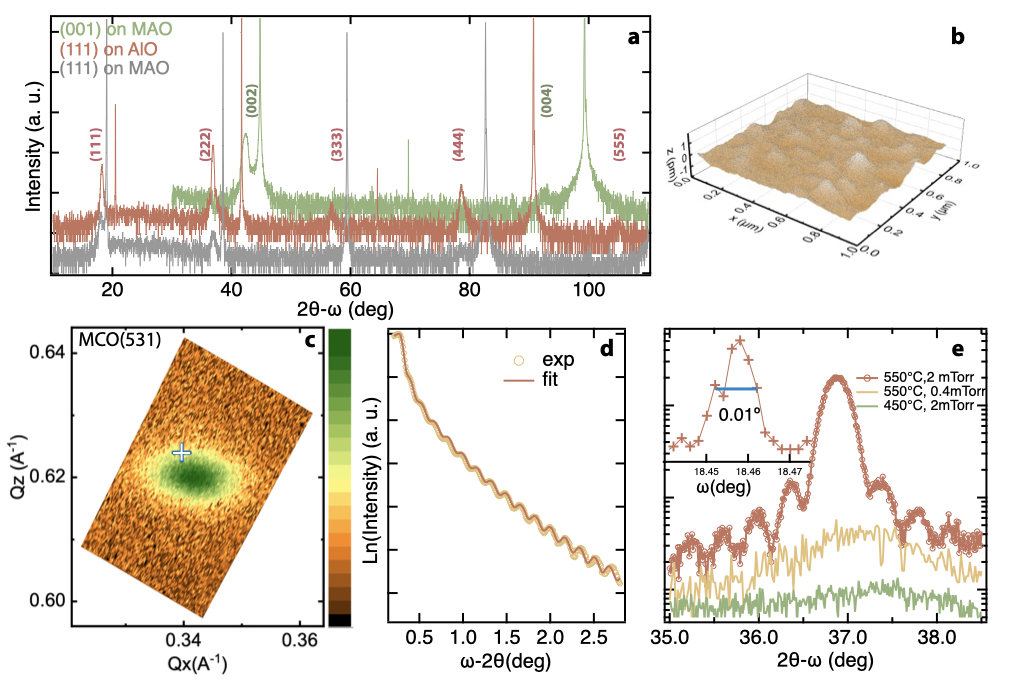}
\caption{\textbf{(a)} Full-range XRD scans of MCO(001) film grown on MAO, MCO(111) film grown on AlO (orange), and MCO(111) film grown on MAO under the same deposition condition. \textbf{(b)} An atomic force microscopy image of MCO(111) grown on the AlO substrate. The averaged RMS roughness in the $1\mu m\times 1\mu m$ area is 171pm. \textbf{(c)} RSM of MCO (531) film peak measured on 80nm MCO (111) film. \textbf{(d)} XRR of the MCO(111) film, whose XRD  is shown in (a), along with the fitting result. From the fitting, the film thickness is estimated to be around 23 nm. \textbf{(e)} Zoomed in XRD scan around  (222) peak measured on films grown under various conditions. Inset is the $\omega$-scan of MCO (111) (550 $^\circ$C, 2mTorr) around  (222) film peak; the estimated FWHM is  0.01$^\circ$.
}
\label{XRD}
\end{figure*}

To further investigate the effect of oxygen pressure and temperature on the growth, a detailed analysis of the crystallinity of a series of samples was carried out by  XRD and XRR.  
Figure \ref{XRD} \textbf{a} shows that no additional chemical phases were observed in any of the films regardless of the orientation or the growth mode. 
Moreover, for similar thicknesses, the MCO(111) on the MAO substrate demonstrates much weaker film peaks than the film grown on AlO, further confirming that AlO substrate with an oxygen-matched sublattice indeed markedly improves the crystallinity of the MCO films. 

In what follows, we focus on the AFM, XRR, and RSM results conducted on the MCO grown on AlO.
The analysis of an XRD scan shown in Fig. \ref{XRD}\textbf{a} yields the out-of-plane lattice constant of  4.87 \AA\,  which under the assumption of a tetragonal distortion results in $\sim$ 0.6\% in-plane elongation which is in accord with the tensile in-plane strain.
XRR data (Fig. \ref{XRD}\textbf{d}) confirm the presence of a very homogeneous film texture with a slow exponential intensity roll-off due to the surface roughness $\sigma  =$ 436 $\pm$ 3 pm. 
As illustrated in Fig. \ref{XRD}
\textbf{b, } this low value of roughness was corroborated by the atomic force microscopy result (RMS roughness of  $\sim$ 171pm). 
Besides, the $\omega$-scan measured around the (222) film Bragg reflection shows a very sharp single peak with FWHM of $\sim$0.01$^\circ$ (see inset of Fig. \ref{XRD}\textbf{e}), further affirming the excellent film crystallinity.
To verify the strain state, we measured reciprocal space map (RSM) around the off-symmetry (531) peak on a thicker sample of 80 nm (see Fig. \ref{XRD} \textbf{c)}. 
By comparing the observed film peak position to the calculated bulk value marked by a white cross, it is clear that the film is compressed in-plane and stretched along the c-axis (i.e., tetragonally distorted). 
The estimated in-plane strain value is -0.5\% which  is in a good agreement with  the value obtained from the $2\theta$ - $\omega$ XRD\ data  (- 0.6\%). 
Overall, these results confirm that imposed strain is robust and is only partially relaxed in the thick MCO(111) film.
In sharp contrast to the (111) oriented growth, for the MCO(100), the out-of-plane lattice constant remained practically unchanged after deposition, which is s consistent with the Stranski-Krastanov growth mode initially revealed by in-situ RHEED. 

In addition to the growth of MCO under the ideal growth condition, other growth conditions have been thoroughly  explored. Here, we  briefly discuss the film crystallinity dependence on substrate, temperature, and pressure on (111)-oriented MCO films. 
Figure \ref{XRD}\textbf{e} shows the MCO (222) Bragg reflection for three representative samples grown under different deposition conditions: S1 - 550 $^\circ$C, 2 mTorr, S2 -  550 $^\circ$C, 0.4 mTorr, and S3 - 450 $^\circ$C, 2mTorr. 
By zooming in to the (222)  reflection, it is clear that there is a distinct oxygen pressure boundary at about 1 mTorr and a temperature threshold at $\sim$ 500 $^\circ$C.  
Below these boundaries, the out-of-plane lattice constant estimated from the peak position becomes closer to the bulk value of 4.83 \AA\ for S2 and 4.81 \AA\ for S3. 
Besides that, the changes in temperature and oxygen pressure do not affect MCO's growth mode along (001).
Finally, all films become amorphous when grown under vacuum below 10$^{-4}$ Torr, as evidenced by the absence of film peaks in XRD.

\begin{figure*}[t]
\vspace{-0pt}
\includegraphics[width=0.8\textwidth]{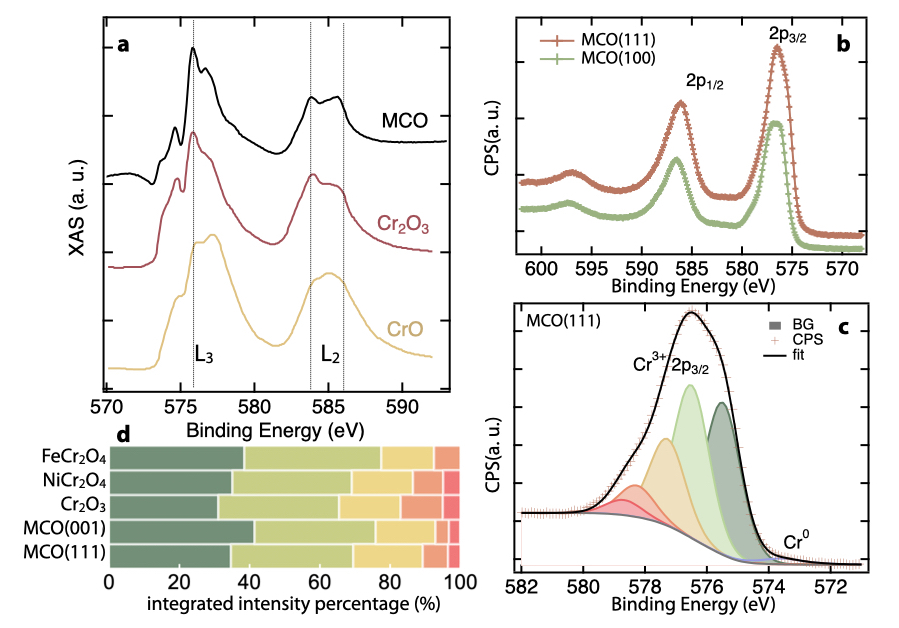}
\caption{\textbf{a.} XAS of MCO (111) measured at room temperature in comparison to  XAS of Cr$_2$O$_3$ and CrO. Cr$_2$O$_3$ and CrO XAS were adapted from REF  \cite{Direct2014HN}. \textbf{b.} Experimental full-range XPS result for MCO (111) and MCO(001). \textbf{c.} Zoomed-in Cr 2p$_{3/2}$ XPS with fitting. The five main peaks are the fine structure of Cr$^{3+}$ as a result of multi-peak splitting, which is standard for chromium (III) oxides. The other small peak is Cr$^0$ due to the surface charging effect. The $\chi^2$ of this fit is 0.399. \textbf{d.} The integrated intensity of the five peaks shown in \textbf{c} in percentage, observed in other closely related systems\cite{X-ray2004MB,X-ray2011BP} compared to the fitting result of MCO(111) and MCO(100).}
\label{XASXPS}
\end{figure*}

Next, we turn our attention to the electronic structure of the films. 
The electronic state and elemental composition were analyzed for MCO films grown in different orientations. 
First, since the X-ray absorption (XAS)\ can fingerprint  the electronic configuration and spin state of MCO, we investigated the films by 
XAS at  Cr L$_{2,3}$-edge at room temperature. 
Figure \ref{XASXPS}\textbf{a} shows the absorption data taken on the MCO films in comparison to the well-known Cr$^{2+}$ and Cr$^{3+}$ reference compounds.
As immediately seen, apart from some minor intensity variations, spectral features of MCO(111) largely mimic those of Cr$_2$O$_3$ (Cr$^{3+}$) and are distinct from the Cr$^{2+}$ spectral line shape.
Additional evidence for expected Cr$^{3+}$  was obtained from the XPS measurement showing the presence of a clear 2\textit{p} doublet, and a pair of shake-up satellite peaks following the doublet at higher binding energy. 
Next, we fitted the chromium 2\textit{p}$_{3/2}$ peak following the Cr$_2$O$_3$ (Cr$^{3+}$)\  references\cite{X-ray2004MB,X-ray2011BP}. 
As shown in Fig. \ref{XASXPS}\textbf{c}, by comparing the MCO(111) peak distribution to the  reference Cr$^{3+}$ compounds CrO, Cr$_2$O$_3$, NiCr$_2$O$_4$ and FeCr$_2$O$_4$, we conclude that the peak positions and their relative intensity indeed closely follow the XPS results for the Cr$^{3+}$ systems with analogous oxygen coordination. 
Interestingly, despite the 3D growth of MCO(100), the overall XPS core-state positions and the fine structure  closely resemble the other chromium Cr$^{3+}$ compounds as well as MCO (111). 
Specifically, no extra peaks, such as Cr IV, V, or VI, have been detected by XPS.

\section{Discussion}

During our search for an ideal growth condition for MCO (001), other substrates (e.g., MgO) and growth conditions (O$_2$ pressure varied from 10$^{-5}$ Torr to 0.1 Torr, the temperature varied from 450$^\circ$C to 750 $^\circ$C) have been explored. 
Despite these efforts, the MCO(001) films form in one of two ways: either develop 3D growth with high roughness, or exhibit poor micro-crystallinity though with a correct charge and chemical state. 
In sharp contrast, MCO(111) films can be grown in a layer-by-layer way, as seen in RHEED and XRD.
Thus it is natural to ask: Why MCO(111)  easier to stabilize than MCO(001)? 

To address this question, we have made a Gibbs free energy comparison according to the model of heteroepitaxial stabilization \cite{Therole2004AK}. 
The energy difference $\Delta E$ after the deposition can be primarily attributed to three parts: $\Delta E=\Delta G_v+ \Delta G_{\textrm{strain}} + \Delta G_{\textrm{surface}}$ where $\Delta G_{\textrm{strain}}\propto \epsilon^2$ ($\epsilon$ is  lattice mismatch), and $\Delta G_v$ is the specific volume Gibbs energy, which  depends only on the initial and final chemical composition. 
For MCO, $\Delta G_v$ is approximately -49200 ($\pm$ 400) J at the growth temperature.\cite{Potentiometric1977KT} 
The negative value of $\Delta E$ means that the film growth is energetically favored.
Now we can compare  MCO (001) and MCO (111) growth based on the energy argument; 
first, experimentally we have determined that MCO (111) stabilizes easier than MCO (001), i.e., $\Delta E^{111}<\Delta E^{001}$. 
Secondly, the XPS result shows that the end composition for both orientations is a pure MCO spinel without secondary chemical phases, therefore $\Delta G_v^{111}=\Delta G_v^{001}$.
Next, for the same substrate but under different orientations, we observed better growth on  MAO (111) substrate than MAO (001). 
Last  but not least, we take in to account the fact that AlO, as a substrate has a larger lattice mismatch $\epsilon $ than MAO (i.e. $\Delta G_{\textrm{strain}}^{111}>\Delta G_{\textrm{strain}}^{001}$).  Finally, 
taking the above three points into consideration, one can conclude that indeed $\Delta G_{\textrm{surface}}^{111}<\Delta G_{\textrm{surface}}^{100}$.  

This result is further corroborated by the data on spinel MAO and other spinel ferrites, showing that (111) surface has significantly lower surface energy than (001) surface. \cite{Surface1977RM} More specifically, the specific surface energy can be written as $\nu = \lambda h^2/\pi^2d$ where $\lambda$ is the elastic modulus normal to the plane under consideration, $d$ is the periodicity, and $h$ is the inter-layer spacing. For different orientations,  $\lambda_{(001)}=C_{11}$ and $\lambda_{(111)}=(1/3)(C_{11}+2C_{12}+4C_{44})$.
By adapting the elastic constants $C_{11}$= 315.53 GPa, $C_{12}$= 206.73 GPa, $C_{44}$= 90.10 GPa \cite{observation2012TW,therole2016SL} one can estimate the surface energy of MCO along the (001) and (111) direction to be $\sim$1664 and 276 erg/cm$^2$, respectively. 
This large disparity in the surface energy strongly influences the initial nucleation process lending strong support to the hypothesis that MCO should preferentially stabilize on the (111) surface, which overall agrees well with the data. 

In conclusion, we have grown the first epitaxially stabilized  MCO(111) films on the Al$_2$O$_3$ substrate. The  MCO films deposited along (111) direction show excellent layer-by-layer growth while MCO films growth along (001) show Stranski-Krastanov  (island-like) growth mode, albeit with correct morphology and chemical composition. 
The samples' high crystallinity was evaluated by XRR and XRD, and the chemical composition and electronic structure confirmed by XPS and XAS.  
The high-quality growth along (111) is explained in the framework of heteroepitaxial stabilization and attributed primarily to the markedly lower surface energy of MCO(111) compared to (001). 
The presented methodology of  growth of (111)-oriented spinel thin films based on the symmetry matching  of  anyon sublattices across the interface  opens up rich research opportunities to probe  spin dynamics in novel frustrated 2D systems.

\section{Acknowledgement}

F. W. was supported by the Claud Lovelace Graduate Fellowship and Department of Energy Grant No. DE-SC0012375. 
X. L., M. K., T.-C. W., M. T. and J. C. acknowledged the support by the Gordon and Betty Moore Foundation EPiQS Initiative through Grant No. GBMF4534. 
This research used resources of the Advanced Light Source, which is a DOE Office of Science User Facility under contract no. DE-AC02-05CH11231.

\bibliography{main}

\begin{thebibliography}{63}%
\makeatletter
\providecommand \@ifxundefined [1]{%
 \@ifx{#1\undefined}
}%
\providecommand \@ifnum [1]{%
 \ifnum #1\expandafter \@firstoftwo
 \else \expandafter \@secondoftwo
 \fi
}%
\providecommand \@ifx [1]{%
 \ifx #1\expandafter \@firstoftwo
 \else \expandafter \@secondoftwo
 \fi
}%
\providecommand \natexlab [1]{#1}%
\providecommand \enquote  [1]{``#1''}%
\providecommand \bibnamefont  [1]{#1}%
\providecommand \bibfnamefont [1]{#1}%
\providecommand \citenamefont [1]{#1}%
\providecommand \href@noop [0]{\@secondoftwo}%
\providecommand \href [0]{\begingroup \@sanitize@url \@href}%
\providecommand \@href[1]{\@@startlink{#1}\@@href}%
\providecommand \@@href[1]{\endgroup#1\@@endlink}%
\providecommand \@sanitize@url [0]{\catcode `\\12\catcode `\$12\catcode
  `\&12\catcode `\#12\catcode `\^12\catcode `\_12\catcode `\%12\relax}%
\providecommand \@@startlink[1]{}%
\providecommand \@@endlink[0]{}%
\providecommand \url  [0]{\begingroup\@sanitize@url \@url }%
\providecommand \@url [1]{\endgroup\@href {#1}{\urlprefix }}%
\providecommand \urlprefix  [0]{URL }%
\providecommand \Eprint [0]{\href }%
\providecommand \doibase [0]{https://doi.org/}%
\providecommand \selectlanguage [0]{\@gobble}%
\providecommand \bibinfo  [0]{\@secondoftwo}%
\providecommand \bibfield  [0]{\@secondoftwo}%
\providecommand \translation [1]{[#1]}%
\providecommand \BibitemOpen [0]{}%
\providecommand \bibitemStop [0]{}%
\providecommand \bibitemNoStop [0]{.\EOS\space}%
\providecommand \EOS [0]{\spacefactor3000\relax}%
\providecommand \BibitemShut  [1]{\csname bibitem#1\endcsname}%
\let\auto@bib@innerbib\@empty
\bibitem [{\citenamefont {Balents}(2010)}]{spin2010LB}%
  \BibitemOpen
  \bibfield  {author} {\bibinfo {author} {\bibfnamefont {L.}~\bibnamefont
  {Balents}},\ }\bibfield  {title} {\bibinfo {title} {Spin liquids in
  frustrated magnets},\ }\href@noop {} {\bibfield  {journal} {\bibinfo
  {journal} {Nature}\ }\textbf {\bibinfo {volume} {464}},\ \bibinfo {pages}
  {199} (\bibinfo {year} {2010})}\BibitemShut {NoStop}%
\bibitem [{\citenamefont {Wen}\ \emph {et~al.}(2019)\citenamefont {Wen},
  \citenamefont {Yu}, \citenamefont {Li}, \citenamefont {Yu},\ and\
  \citenamefont {Li}}]{experimental2019JW}%
  \BibitemOpen
  \bibfield  {author} {\bibinfo {author} {\bibfnamefont {J.~S.}\ \bibnamefont
  {Wen}}, \bibinfo {author} {\bibfnamefont {S.~L.}\ \bibnamefont {Yu}},
  \bibinfo {author} {\bibfnamefont {S.~Y.}\ \bibnamefont {Li}}, \bibinfo
  {author} {\bibfnamefont {W.~Q.}\ \bibnamefont {Yu}},\ and\ \bibinfo {author}
  {\bibfnamefont {J.~X.}\ \bibnamefont {Li}},\ }\bibfield  {title} {\bibinfo
  {title} {Experimental identification of quantum spin liquids},\ }\href {<Go
  to ISI>://WOS:000470239900001} {\bibfield  {journal} {\bibinfo  {journal}
  {Npj Quantum Materials}\ }\textbf {\bibinfo {volume} {4}},\ \bibinfo {pages}
  {9} (\bibinfo {year} {2019})}\BibitemShut {NoStop}%
\bibitem [{\citenamefont {Zhou}\ \emph {et~al.}(2017)\citenamefont {Zhou},
  \citenamefont {Kanoda},\ and\ \citenamefont {Ng}}]{quantum2017YZ}%
  \BibitemOpen
  \bibfield  {author} {\bibinfo {author} {\bibfnamefont {Y.}~\bibnamefont
  {Zhou}}, \bibinfo {author} {\bibfnamefont {K.}~\bibnamefont {Kanoda}},\ and\
  \bibinfo {author} {\bibfnamefont {T.~K.}\ \bibnamefont {Ng}},\ }\bibfield
  {title} {\bibinfo {title} {Quantum spin liquid states},\ }\href {<Go to
  ISI>://WOS:000399385200001} {\bibfield  {journal} {\bibinfo  {journal}
  {Reviews of Modern Physics}\ }\textbf {\bibinfo {volume} {89}},\ \bibinfo
  {pages} {50} (\bibinfo {year} {2017})}\BibitemShut {NoStop}%
\bibitem [{\citenamefont {Norman}(2016)}]{Herbertsmithite2016MN}%
  \BibitemOpen
  \bibfield  {author} {\bibinfo {author} {\bibfnamefont {M.~R.}\ \bibnamefont
  {Norman}},\ }\bibfield  {title} {\bibinfo {title} {Colloquium:
  Herbertsmithite and the search for the quantum spin liquid},\ }\href {<Go to
  ISI>://WOS:000389033500001} {\bibfield  {journal} {\bibinfo  {journal}
  {Reviews of Modern Physics}\ }\textbf {\bibinfo {volume} {88}},\ \bibinfo
  {pages} {14} (\bibinfo {year} {2016})}\BibitemShut {NoStop}%
\bibitem [{\citenamefont {Zhu}\ \emph {et~al.}(2020)\citenamefont {Zhu},
  \citenamefont {Kong}, \citenamefont {Cao}, \citenamefont {Chen},
  \citenamefont {Papaj}, \citenamefont {Du}, \citenamefont {Xing},
  \citenamefont {Liu}, \citenamefont {Wang}, \citenamefont {Shen},
  \citenamefont {Yang}, \citenamefont {Schneeloch}, \citenamefont {Zhong},
  \citenamefont {Gu}, \citenamefont {Fu}, \citenamefont {Zhang}, \citenamefont
  {Ding},\ and\ \citenamefont {Gao}}]{Nearly2020SZ}%
  \BibitemOpen
  \bibfield  {author} {\bibinfo {author} {\bibfnamefont {S.~Y.}\ \bibnamefont
  {Zhu}}, \bibinfo {author} {\bibfnamefont {L.~Y.}\ \bibnamefont {Kong}},
  \bibinfo {author} {\bibfnamefont {L.}~\bibnamefont {Cao}}, \bibinfo {author}
  {\bibfnamefont {H.}~\bibnamefont {Chen}}, \bibinfo {author} {\bibfnamefont
  {M.}~\bibnamefont {Papaj}}, \bibinfo {author} {\bibfnamefont {S.~X.}\
  \bibnamefont {Du}}, \bibinfo {author} {\bibfnamefont {Y.~Q.}\ \bibnamefont
  {Xing}}, \bibinfo {author} {\bibfnamefont {W.~Y.}\ \bibnamefont {Liu}},
  \bibinfo {author} {\bibfnamefont {D.~F.}\ \bibnamefont {Wang}}, \bibinfo
  {author} {\bibfnamefont {C.~M.}\ \bibnamefont {Shen}}, \bibinfo {author}
  {\bibfnamefont {F.~Z.}\ \bibnamefont {Yang}}, \bibinfo {author}
  {\bibfnamefont {J.}~\bibnamefont {Schneeloch}}, \bibinfo {author}
  {\bibfnamefont {R.~D.}\ \bibnamefont {Zhong}}, \bibinfo {author}
  {\bibfnamefont {G.~D.}\ \bibnamefont {Gu}}, \bibinfo {author} {\bibfnamefont
  {L.}~\bibnamefont {Fu}}, \bibinfo {author} {\bibfnamefont {Y.~Y.}\
  \bibnamefont {Zhang}}, \bibinfo {author} {\bibfnamefont {H.}~\bibnamefont
  {Ding}},\ and\ \bibinfo {author} {\bibfnamefont {H.~J.}\ \bibnamefont
  {Gao}},\ }\bibfield  {title} {\bibinfo {title} {Nearly quantized conductance
  plateau of vortex zero mode in an iron-based superconductor},\ }\href {<Go to
  ISI>://WOS:000506811300039} {\bibfield  {journal} {\bibinfo  {journal}
  {Science}\ }\textbf {\bibinfo {volume} {367}},\ \bibinfo {pages} {189}
  (\bibinfo {year} {2020})}\BibitemShut {NoStop}%
\bibitem [{\citenamefont {Koborinai}\ \emph {et~al.}(2016)\citenamefont
  {Koborinai}, \citenamefont {Dissanayake}, \citenamefont {Reehuis},
  \citenamefont {Matsuda}, \citenamefont {Kajita}, \citenamefont {Kuwahara},
  \citenamefont {Lee},\ and\ \citenamefont {Katsufuji}}]{Orbital2016RK}%
  \BibitemOpen
  \bibfield  {author} {\bibinfo {author} {\bibfnamefont {R.}~\bibnamefont
  {Koborinai}}, \bibinfo {author} {\bibfnamefont {S.~E.}\ \bibnamefont
  {Dissanayake}}, \bibinfo {author} {\bibfnamefont {M.}~\bibnamefont
  {Reehuis}}, \bibinfo {author} {\bibfnamefont {M.}~\bibnamefont {Matsuda}},
  \bibinfo {author} {\bibfnamefont {T.}~\bibnamefont {Kajita}}, \bibinfo
  {author} {\bibfnamefont {H.}~\bibnamefont {Kuwahara}}, \bibinfo {author}
  {\bibfnamefont {S.~H.}\ \bibnamefont {Lee}},\ and\ \bibinfo {author}
  {\bibfnamefont {T.}~\bibnamefont {Katsufuji}},\ }\bibfield  {title} {\bibinfo
  {title} {Orbital glass state of the nearly metallic spinel cobalt vanadate},\
  }\href {<Go to ISI>://WOS:000368523200011} {\bibfield  {journal} {\bibinfo
  {journal} {Physical Review Letters}\ }\textbf {\bibinfo {volume} {116}}
  (\bibinfo {year} {2016})}\BibitemShut {NoStop}%
\bibitem [{\citenamefont {Uehara}\ \emph {et~al.}(2015)\citenamefont {Uehara},
  \citenamefont {Shinaoka},\ and\ \citenamefont {Motome}}]{Charge2015AU}%
  \BibitemOpen
  \bibfield  {author} {\bibinfo {author} {\bibfnamefont {A.}~\bibnamefont
  {Uehara}}, \bibinfo {author} {\bibfnamefont {H.}~\bibnamefont {Shinaoka}},\
  and\ \bibinfo {author} {\bibfnamefont {Y.}~\bibnamefont {Motome}},\
  }\bibfield  {title} {\bibinfo {title} {Charge-spin-orbital fluctuations in
  mixed valence spinels: Comparative study of alv2o4 and liv2o4},\ }\href {<Go
  to ISI>://WOS:000365773300001} {\bibfield  {journal} {\bibinfo  {journal}
  {Physical Review B}\ }\textbf {\bibinfo {volume} {92}} (\bibinfo {year}
  {2015})}\BibitemShut {NoStop}%
\bibitem [{\citenamefont {Avella}\ \emph {et~al.}(2019)\citenamefont {Avella},
  \citenamefont {Oles},\ and\ \citenamefont {Horsch}}]{Defect2019AA}%
  \BibitemOpen
  \bibfield  {author} {\bibinfo {author} {\bibfnamefont {A.}~\bibnamefont
  {Avella}}, \bibinfo {author} {\bibfnamefont {A.~M.}\ \bibnamefont {Oles}},\
  and\ \bibinfo {author} {\bibfnamefont {P.}~\bibnamefont {Horsch}},\
  }\bibfield  {title} {\bibinfo {title} {Defect-induced orbital polarization
  and collapse of orbital order in doped vanadium perovskites},\ }\href {<Go to
  ISI>://WOS:000462936100013} {\bibfield  {journal} {\bibinfo  {journal}
  {Physical Review Letters}\ }\textbf {\bibinfo {volume} {122}} (\bibinfo
  {year} {2019})}\BibitemShut {NoStop}%
\bibitem [{\citenamefont {Shimizu}\ \emph {et~al.}(2007)\citenamefont
  {Shimizu}, \citenamefont {Akimoto}, \citenamefont {Tsujii}, \citenamefont
  {Tajima},\ and\ \citenamefont {Kato}}]{Mott2007YS}%
  \BibitemOpen
  \bibfield  {author} {\bibinfo {author} {\bibfnamefont {Y.}~\bibnamefont
  {Shimizu}}, \bibinfo {author} {\bibfnamefont {H.}~\bibnamefont {Akimoto}},
  \bibinfo {author} {\bibfnamefont {H.}~\bibnamefont {Tsujii}}, \bibinfo
  {author} {\bibfnamefont {A.}~\bibnamefont {Tajima}},\ and\ \bibinfo {author}
  {\bibfnamefont {R.}~\bibnamefont {Kato}},\ }\bibfield  {title} {\bibinfo
  {title} {Mott transition in a valence-bond solid insulator with a triangular
  lattice},\ }\href {<Go to ISI>://WOS:000251887100039} {\bibfield  {journal}
  {\bibinfo  {journal} {Physical Review Letters}\ }\textbf {\bibinfo {volume}
  {99}} (\bibinfo {year} {2007})}\BibitemShut {NoStop}%
\bibitem [{\citenamefont {Tamura}\ \emph {et~al.}(2006)\citenamefont {Tamura},
  \citenamefont {Nakao},\ and\ \citenamefont {Kato}}]{Frustration2006MT}%
  \BibitemOpen
  \bibfield  {author} {\bibinfo {author} {\bibfnamefont {M.}~\bibnamefont
  {Tamura}}, \bibinfo {author} {\bibfnamefont {A.}~\bibnamefont {Nakao}},\ and\
  \bibinfo {author} {\bibfnamefont {R.}~\bibnamefont {Kato}},\ }\bibfield
  {title} {\bibinfo {title} {Frustration-induced valence-bond ordering in a new
  quantum triangular antiferromagnet based on pd(dmit)(2)},\ }\href {<Go to
  ISI>://WOS:000241809500001} {\bibfield  {journal} {\bibinfo  {journal}
  {Journal of the Physical Society of Japan}\ }\textbf {\bibinfo {volume} {75}}
  (\bibinfo {year} {2006})}\BibitemShut {NoStop}%
\bibitem [{\citenamefont {Schroder}\ \emph {et~al.}(2005)\citenamefont
  {Schroder}, \citenamefont {Nojiri}, \citenamefont {Schnack}, \citenamefont
  {Hage}, \citenamefont {Luban},\ and\ \citenamefont
  {Kogerler}}]{Competing2005CS}%
  \BibitemOpen
  \bibfield  {author} {\bibinfo {author} {\bibfnamefont {C.}~\bibnamefont
  {Schroder}}, \bibinfo {author} {\bibfnamefont {H.}~\bibnamefont {Nojiri}},
  \bibinfo {author} {\bibfnamefont {J.}~\bibnamefont {Schnack}}, \bibinfo
  {author} {\bibfnamefont {P.}~\bibnamefont {Hage}}, \bibinfo {author}
  {\bibfnamefont {M.}~\bibnamefont {Luban}},\ and\ \bibinfo {author}
  {\bibfnamefont {P.}~\bibnamefont {Kogerler}},\ }\bibfield  {title} {\bibinfo
  {title} {Competing spin phases in geometrically frustrated magnetic
  molecules},\ }\href {<Go to ISI>://WOS:000226308000079} {\bibfield  {journal}
  {\bibinfo  {journal} {Physical Review Letters}\ }\textbf {\bibinfo {volume}
  {94}} (\bibinfo {year} {2005})}\BibitemShut {NoStop}%
\bibitem [{\citenamefont {Baker}\ \emph {et~al.}(2012)\citenamefont {Baker},
  \citenamefont {Timco}, \citenamefont {Piligkos}, \citenamefont {Mathieson},
  \citenamefont {Mutka}, \citenamefont {Tuna}, \citenamefont {Kozlowski},
  \citenamefont {Antkowiak}, \citenamefont {Guidi}, \citenamefont {Gupta},
  \citenamefont {Rath}, \citenamefont {Woolfson}, \citenamefont {Kamieniarz},
  \citenamefont {Pritchard}, \citenamefont {Weihe}, \citenamefont {Cronin},
  \citenamefont {Rajaraman}, \citenamefont {Collison}, \citenamefont
  {McInnes},\ and\ \citenamefont {Winpenny}}]{Classification2012MB}%
  \BibitemOpen
  \bibfield  {author} {\bibinfo {author} {\bibfnamefont {M.~L.}\ \bibnamefont
  {Baker}}, \bibinfo {author} {\bibfnamefont {G.~A.}\ \bibnamefont {Timco}},
  \bibinfo {author} {\bibfnamefont {S.}~\bibnamefont {Piligkos}}, \bibinfo
  {author} {\bibfnamefont {J.~S.}\ \bibnamefont {Mathieson}}, \bibinfo {author}
  {\bibfnamefont {H.}~\bibnamefont {Mutka}}, \bibinfo {author} {\bibfnamefont
  {F.}~\bibnamefont {Tuna}}, \bibinfo {author} {\bibfnamefont {P.}~\bibnamefont
  {Kozlowski}}, \bibinfo {author} {\bibfnamefont {M.}~\bibnamefont
  {Antkowiak}}, \bibinfo {author} {\bibfnamefont {T.}~\bibnamefont {Guidi}},
  \bibinfo {author} {\bibfnamefont {T.}~\bibnamefont {Gupta}}, \bibinfo
  {author} {\bibfnamefont {H.}~\bibnamefont {Rath}}, \bibinfo {author}
  {\bibfnamefont {R.~J.}\ \bibnamefont {Woolfson}}, \bibinfo {author}
  {\bibfnamefont {G.}~\bibnamefont {Kamieniarz}}, \bibinfo {author}
  {\bibfnamefont {R.~G.}\ \bibnamefont {Pritchard}}, \bibinfo {author}
  {\bibfnamefont {H.}~\bibnamefont {Weihe}}, \bibinfo {author} {\bibfnamefont
  {L.}~\bibnamefont {Cronin}}, \bibinfo {author} {\bibfnamefont
  {G.}~\bibnamefont {Rajaraman}}, \bibinfo {author} {\bibfnamefont
  {D.}~\bibnamefont {Collison}}, \bibinfo {author} {\bibfnamefont {E.~J.~L.}\
  \bibnamefont {McInnes}},\ and\ \bibinfo {author} {\bibfnamefont {R.~E.~P.}\
  \bibnamefont {Winpenny}},\ }\bibfield  {title} {\bibinfo {title} {A
  classification of spin frustration in molecular magnets from a physical study
  of large odd-numbered-metal, odd electron rings},\ }\href {<Go to
  ISI>://WOS:000311997200027} {\bibfield  {journal} {\bibinfo  {journal}
  {Proceedings of the National Academy of Sciences of the United States of
  America}\ }\textbf {\bibinfo {volume} {109}},\ \bibinfo {pages} {19113}
  (\bibinfo {year} {2012})}\BibitemShut {NoStop}%
\bibitem [{\citenamefont {Garlea}\ \emph {et~al.}(2019)\citenamefont {Garlea},
  \citenamefont {Sanjeewa}, \citenamefont {McGuire}, \citenamefont {Batista},
  \citenamefont {Samarakoon}, \citenamefont {Graf}, \citenamefont {Winn},
  \citenamefont {Ye}, \citenamefont {Hoffmann},\ and\ \citenamefont
  {Kolis}}]{Exotic2019VOG}%
  \BibitemOpen
  \bibfield  {author} {\bibinfo {author} {\bibfnamefont {V.~O.}\ \bibnamefont
  {Garlea}}, \bibinfo {author} {\bibfnamefont {L.~D.}\ \bibnamefont
  {Sanjeewa}}, \bibinfo {author} {\bibfnamefont {M.~A.}\ \bibnamefont
  {McGuire}}, \bibinfo {author} {\bibfnamefont {C.~D.}\ \bibnamefont
  {Batista}}, \bibinfo {author} {\bibfnamefont {A.~M.}\ \bibnamefont
  {Samarakoon}}, \bibinfo {author} {\bibfnamefont {D.}~\bibnamefont {Graf}},
  \bibinfo {author} {\bibfnamefont {B.}~\bibnamefont {Winn}}, \bibinfo {author}
  {\bibfnamefont {F.}~\bibnamefont {Ye}}, \bibinfo {author} {\bibfnamefont
  {C.}~\bibnamefont {Hoffmann}},\ and\ \bibinfo {author} {\bibfnamefont
  {J.~W.}\ \bibnamefont {Kolis}},\ }\bibfield  {title} {\bibinfo {title}
  {Exotic magnetic field-induced spin-superstructures in a mixed
  honeycomb-triangular lattice system},\ }\href {<Go to
  ISI>://WOS:000459918600001} {\bibfield  {journal} {\bibinfo  {journal}
  {Physical Review X}\ }\textbf {\bibinfo {volume} {9}} (\bibinfo {year}
  {2019})}\BibitemShut {NoStop}%
\bibitem [{\citenamefont {Savary}\ and\ \citenamefont
  {Balents}(2017)}]{Quantum2016LS}%
  \BibitemOpen
  \bibfield  {author} {\bibinfo {author} {\bibfnamefont {L.}~\bibnamefont
  {Savary}}\ and\ \bibinfo {author} {\bibfnamefont {L.}~\bibnamefont
  {Balents}},\ }\bibfield  {title} {\bibinfo {title} {Quantum spin liquids: a
  review},\ }\href {<Go to ISI>://WOS:000388231300001} {\bibfield  {journal}
  {\bibinfo  {journal} {Reports on Progress in Physics}\ }\textbf {\bibinfo
  {volume} {80}} (\bibinfo {year} {2017})}\BibitemShut {NoStop}%
\bibitem [{\citenamefont {Hermanns}\ \emph {et~al.}(2018)\citenamefont
  {Hermanns}, \citenamefont {Kimchi},\ and\ \citenamefont
  {Knolle}}]{Physics2018MH}%
  \BibitemOpen
  \bibfield  {author} {\bibinfo {author} {\bibfnamefont {M.}~\bibnamefont
  {Hermanns}}, \bibinfo {author} {\bibfnamefont {I.}~\bibnamefont {Kimchi}},\
  and\ \bibinfo {author} {\bibfnamefont {J.}~\bibnamefont {Knolle}},\ }\bibinfo
  {title} {Physics of the kitaev model: Fractionalization, dynamic
  correlations, and material connections},\ in\ \href {<Go to
  ISI>://WOS:000429191300002} {\emph {\bibinfo {booktitle} {Annual Review of
  Condensed Matter Physics}}},\ Vol.~\bibinfo {volume} {9},\ \bibinfo {editor}
  {edited by\ \bibinfo {editor} {\bibfnamefont {S.}~\bibnamefont {Sachdev}}\
  and\ \bibinfo {editor} {\bibfnamefont {M.~C.}\ \bibnamefont {Marchetti}}}\
  (\bibinfo {year} {2018})\ pp.\ \bibinfo {pages} {17--33}\BibitemShut
  {NoStop}%
\bibitem [{\citenamefont {Takagi}\ \emph {et~al.}(2019)\citenamefont {Takagi},
  \citenamefont {Takayama}, \citenamefont {Jackeli}, \citenamefont
  {Khaliullin},\ and\ \citenamefont {Nagler}}]{Concept2019HT}%
  \BibitemOpen
  \bibfield  {author} {\bibinfo {author} {\bibfnamefont {H.}~\bibnamefont
  {Takagi}}, \bibinfo {author} {\bibfnamefont {T.}~\bibnamefont {Takayama}},
  \bibinfo {author} {\bibfnamefont {G.}~\bibnamefont {Jackeli}}, \bibinfo
  {author} {\bibfnamefont {G.}~\bibnamefont {Khaliullin}},\ and\ \bibinfo
  {author} {\bibfnamefont {S.~E.}\ \bibnamefont {Nagler}},\ }\bibfield  {title}
  {\bibinfo {title} {Concept and realization of kitaev quantum spin liquids},\
  }\href {<Go to ISI>://WOS:000540317900009} {\bibfield  {journal} {\bibinfo
  {journal} {Nature Reviews Physics}\ }\textbf {\bibinfo {volume} {1}},\
  \bibinfo {pages} {264} (\bibinfo {year} {2019})}\BibitemShut {NoStop}%
\bibitem [{\citenamefont {Broholm}\ \emph {et~al.}(2020)\citenamefont
  {Broholm}, \citenamefont {Cava}, \citenamefont {Kivelson}, \citenamefont
  {Nocera}, \citenamefont {Norman},\ and\ \citenamefont
  {Senthil}}]{Quantum2020CB}%
  \BibitemOpen
  \bibfield  {author} {\bibinfo {author} {\bibfnamefont {C.}~\bibnamefont
  {Broholm}}, \bibinfo {author} {\bibfnamefont {R.~J.}\ \bibnamefont {Cava}},
  \bibinfo {author} {\bibfnamefont {S.~A.}\ \bibnamefont {Kivelson}}, \bibinfo
  {author} {\bibfnamefont {D.~G.}\ \bibnamefont {Nocera}}, \bibinfo {author}
  {\bibfnamefont {M.~R.}\ \bibnamefont {Norman}},\ and\ \bibinfo {author}
  {\bibfnamefont {T.}~\bibnamefont {Senthil}},\ }\bibfield  {title} {\bibinfo
  {title} {Quantum spin liquids},\ }\href {<Go to ISI>://WOS:000509802700029}
  {\bibfield  {journal} {\bibinfo  {journal} {Science}\ }\textbf {\bibinfo
  {volume} {367}},\ \bibinfo {pages} {263} (\bibinfo {year}
  {2020})}\BibitemShut {NoStop}%
\bibitem [{\citenamefont {Gardner}\ \emph {et~al.}(2010)\citenamefont
  {Gardner}, \citenamefont {Gingras},\ and\ \citenamefont
  {Greedan}}]{magnetic2010JG}%
  \BibitemOpen
  \bibfield  {author} {\bibinfo {author} {\bibfnamefont {J.~S.}\ \bibnamefont
  {Gardner}}, \bibinfo {author} {\bibfnamefont {M.~J.~P.}\ \bibnamefont
  {Gingras}},\ and\ \bibinfo {author} {\bibfnamefont {J.~E.}\ \bibnamefont
  {Greedan}},\ }\bibfield  {title} {\bibinfo {title} {Magnetic pyrochlore
  oxides},\ }\href {<Go to ISI>://WOS:000276184000002} {\bibfield  {journal}
  {\bibinfo  {journal} {Reviews of Modern Physics}\ }\textbf {\bibinfo {volume}
  {82}},\ \bibinfo {pages} {53} (\bibinfo {year} {2010})}\BibitemShut {NoStop}%
\bibitem [{\citenamefont {Rau}\ and\ \citenamefont
  {Gingras}(2019)}]{Frustrated2019JR}%
  \BibitemOpen
  \bibfield  {author} {\bibinfo {author} {\bibfnamefont {J.~G.}\ \bibnamefont
  {Rau}}\ and\ \bibinfo {author} {\bibfnamefont {M.~J.~P.}\ \bibnamefont
  {Gingras}},\ }\bibinfo {title} {Frustrated quantum rare-earth pyrochlores},\
  in\ \href {<Go to ISI>://WOS:000461414200018} {\emph {\bibinfo {booktitle}
  {Annual Review of Condensed Matter Physics}}},\ Vol.~\bibinfo {volume} {10},\
  \bibinfo {editor} {edited by\ \bibinfo {editor} {\bibfnamefont
  {S.}~\bibnamefont {Sachdev}}\ and\ \bibinfo {editor} {\bibfnamefont {M.~C.}\
  \bibnamefont {Marchetti}}}\ (\bibinfo  {publisher} {Annual Reviews},\
  \bibinfo {address} {Palo Alto},\ \bibinfo {year} {2019})\ pp.\ \bibinfo
  {pages} {357--386}\BibitemShut {NoStop}%
\bibitem [{\citenamefont {Bramwell}\ and\ \citenamefont
  {Gingras}(2001)}]{Spin2001SB}%
  \BibitemOpen
  \bibfield  {author} {\bibinfo {author} {\bibfnamefont {S.~T.}\ \bibnamefont
  {Bramwell}}\ and\ \bibinfo {author} {\bibfnamefont {M.~J.~P.}\ \bibnamefont
  {Gingras}},\ }\bibfield  {title} {\bibinfo {title} {Spin ice state in
  frustrated magnetic pyrochlore materials},\ }\href {<Go to
  ISI>://WOS:000172240500039} {\bibfield  {journal} {\bibinfo  {journal}
  {Science}\ }\textbf {\bibinfo {volume} {294}},\ \bibinfo {pages} {1495}
  (\bibinfo {year} {2001})}\BibitemShut {NoStop}%
\bibitem [{\citenamefont {Nisoli}\ \emph {et~al.}(2013)\citenamefont {Nisoli},
  \citenamefont {Moessner},\ and\ \citenamefont {Schiffer}}]{Artificial2013CN}%
  \BibitemOpen
  \bibfield  {author} {\bibinfo {author} {\bibfnamefont {C.}~\bibnamefont
  {Nisoli}}, \bibinfo {author} {\bibfnamefont {R.}~\bibnamefont {Moessner}},\
  and\ \bibinfo {author} {\bibfnamefont {P.}~\bibnamefont {Schiffer}},\
  }\bibfield  {title} {\bibinfo {title} {Colloquium: Artificial spin ice:
  Designing and imaging magnetic frustration},\ }\href {<Go to
  ISI>://WOS:000325355200004} {\bibfield  {journal} {\bibinfo  {journal}
  {Reviews of Modern Physics}\ }\textbf {\bibinfo {volume} {85}},\ \bibinfo
  {pages} {1473} (\bibinfo {year} {2013})}\BibitemShut {NoStop}%
\bibitem [{\citenamefont {Castelnovo}\ \emph {et~al.}(2012)\citenamefont
  {Castelnovo}, \citenamefont {Moessner},\ and\ \citenamefont
  {Sondhi}}]{Spin2019CC}%
  \BibitemOpen
  \bibfield  {author} {\bibinfo {author} {\bibfnamefont {C.}~\bibnamefont
  {Castelnovo}}, \bibinfo {author} {\bibfnamefont {R.}~\bibnamefont
  {Moessner}},\ and\ \bibinfo {author} {\bibfnamefont {S.~L.}\ \bibnamefont
  {Sondhi}},\ }\bibinfo {title} {Spin ice, fractionalization, and topological
  order},\ in\ \href {<Go to ISI>://WOS:000301793100004} {\emph {\bibinfo
  {booktitle} {Annual Review of Condensed Matter Physics, Vol 3}}},\ \bibinfo
  {series} {Annual Review of Condensed Matter Physics}, Vol.~\bibinfo {volume}
  {3},\ \bibinfo {editor} {edited by\ \bibinfo {editor} {\bibfnamefont {J.~S.}\
  \bibnamefont {Langer}}}\ (\bibinfo  {publisher} {Annual Reviews},\ \bibinfo
  {address} {Palo Alto},\ \bibinfo {year} {2012})\ pp.\ \bibinfo {pages}
  {35--55}\BibitemShut {NoStop}%
\bibitem [{\citenamefont {Lee}\ \emph {et~al.}(2010)\citenamefont {Lee},
  \citenamefont {Takagi}, \citenamefont {Louca}, \citenamefont {Matsuda},
  \citenamefont {Ji}, \citenamefont {Ueda}, \citenamefont {Ueda}, \citenamefont
  {Katsufuji}, \citenamefont {Chung}, \citenamefont {Park}, \citenamefont
  {Cheong},\ and\ \citenamefont {Broholm}}]{Frustrated2010SL}%
  \BibitemOpen
  \bibfield  {author} {\bibinfo {author} {\bibfnamefont {S.~H.}\ \bibnamefont
  {Lee}}, \bibinfo {author} {\bibfnamefont {H.}~\bibnamefont {Takagi}},
  \bibinfo {author} {\bibfnamefont {D.}~\bibnamefont {Louca}}, \bibinfo
  {author} {\bibfnamefont {M.}~\bibnamefont {Matsuda}}, \bibinfo {author}
  {\bibfnamefont {S.}~\bibnamefont {Ji}}, \bibinfo {author} {\bibfnamefont
  {H.}~\bibnamefont {Ueda}}, \bibinfo {author} {\bibfnamefont {Y.}~\bibnamefont
  {Ueda}}, \bibinfo {author} {\bibfnamefont {T.}~\bibnamefont {Katsufuji}},
  \bibinfo {author} {\bibfnamefont {J.~H.}\ \bibnamefont {Chung}}, \bibinfo
  {author} {\bibfnamefont {S.}~\bibnamefont {Park}}, \bibinfo {author}
  {\bibfnamefont {S.~W.}\ \bibnamefont {Cheong}},\ and\ \bibinfo {author}
  {\bibfnamefont {C.}~\bibnamefont {Broholm}},\ }\bibfield  {title} {\bibinfo
  {title} {Frustrated magnetism and cooperative phase transitions in spinels},\
  }\href {<Go to ISI>://WOS:000273801200005} {\bibfield  {journal} {\bibinfo
  {journal} {Journal of the Physical Society of Japan}\ }\textbf {\bibinfo
  {volume} {79}} (\bibinfo {year} {2010})}\BibitemShut {NoStop}%
\bibitem [{\citenamefont {Yafet}\ and\ \citenamefont
  {Kittel}(1952)}]{Antiferromagnetic1952YY}%
  \BibitemOpen
  \bibfield  {author} {\bibinfo {author} {\bibfnamefont {Y.}~\bibnamefont
  {Yafet}}\ and\ \bibinfo {author} {\bibfnamefont {C.}~\bibnamefont {Kittel}},\
  }\bibfield  {title} {\bibinfo {title} {Antiferromagnetic arrangements in
  ferrites},\ }\href {<Go to ISI>://WOS:A1952UB43700008} {\bibfield  {journal}
  {\bibinfo  {journal} {Physical Review}\ }\textbf {\bibinfo {volume} {87}},\
  \bibinfo {pages} {290} (\bibinfo {year} {1952})}\BibitemShut {NoStop}%
\bibitem [{\citenamefont {Lee}\ \emph {et~al.}(2017)\citenamefont {Lee},
  \citenamefont {Ma}, \citenamefont {Hahn}, \citenamefont {Cao}, \citenamefont
  {Lee}, \citenamefont {Hong}, \citenamefont {Lee}, \citenamefont {Yeom},
  \citenamefont {Okamoto}, \citenamefont {Zhou}, \citenamefont {Matsuda},\ and\
  \citenamefont {Fishman}}]{Magnetic2017JL}%
  \BibitemOpen
  \bibfield  {author} {\bibinfo {author} {\bibfnamefont {J.~H.}\ \bibnamefont
  {Lee}}, \bibinfo {author} {\bibfnamefont {J.}~\bibnamefont {Ma}}, \bibinfo
  {author} {\bibfnamefont {S.~E.}\ \bibnamefont {Hahn}}, \bibinfo {author}
  {\bibfnamefont {H.~B.}\ \bibnamefont {Cao}}, \bibinfo {author} {\bibfnamefont
  {M.}~\bibnamefont {Lee}}, \bibinfo {author} {\bibfnamefont {T.}~\bibnamefont
  {Hong}}, \bibinfo {author} {\bibfnamefont {H.~J.}\ \bibnamefont {Lee}},
  \bibinfo {author} {\bibfnamefont {M.~S.}\ \bibnamefont {Yeom}}, \bibinfo
  {author} {\bibfnamefont {S.}~\bibnamefont {Okamoto}}, \bibinfo {author}
  {\bibfnamefont {H.~D.}\ \bibnamefont {Zhou}}, \bibinfo {author}
  {\bibfnamefont {M.}~\bibnamefont {Matsuda}},\ and\ \bibinfo {author}
  {\bibfnamefont {R.~S.}\ \bibnamefont {Fishman}},\ }\bibfield  {title}
  {\bibinfo {title} {Magnetic frustration driven by itinerancy in spinel
  cov2o4},\ }\href {<Go to ISI>://WOS:000417353600023} {\bibfield  {journal}
  {\bibinfo  {journal} {Scientific Reports}\ }\textbf {\bibinfo {volume} {7}}
  (\bibinfo {year} {2017})}\BibitemShut {NoStop}%
\bibitem [{\citenamefont {Reig-i Plessis}\ \emph {et~al.}(2019)\citenamefont
  {Reig-i Plessis}, \citenamefont {Geldern}, \citenamefont {Aczel},
  \citenamefont {Kochkov}, \citenamefont {Clark},\ and\ \citenamefont
  {MacDougall}}]{Deviation2019DR}%
  \BibitemOpen
  \bibfield  {author} {\bibinfo {author} {\bibfnamefont {D.}~\bibnamefont
  {Reig-i Plessis}}, \bibinfo {author} {\bibfnamefont {S.~V.}\ \bibnamefont
  {Geldern}}, \bibinfo {author} {\bibfnamefont {A.~A.}\ \bibnamefont {Aczel}},
  \bibinfo {author} {\bibfnamefont {D.}~\bibnamefont {Kochkov}}, \bibinfo
  {author} {\bibfnamefont {B.~K.}\ \bibnamefont {Clark}},\ and\ \bibinfo
  {author} {\bibfnamefont {G.~J.}\ \bibnamefont {MacDougall}},\ }\bibfield
  {title} {\bibinfo {title} {Deviation from the dipole-ice model in the spinel
  spin-ice candidate mger2se4},\ }\href {<Go to ISI>://WOS:000466383000002}
  {\bibfield  {journal} {\bibinfo  {journal} {Physical Review B}\ }\textbf
  {\bibinfo {volume} {99}} (\bibinfo {year} {2019})}\BibitemShut {NoStop}%
\bibitem [{\citenamefont {Tristan}\ \emph {et~al.}(2005)\citenamefont
  {Tristan}, \citenamefont {Hemberger}, \citenamefont {Krimmel}, \citenamefont
  {von Nidda}, \citenamefont {Tsurkan},\ and\ \citenamefont
  {Loidl}}]{Geometric2005NT}%
  \BibitemOpen
  \bibfield  {author} {\bibinfo {author} {\bibfnamefont {N.}~\bibnamefont
  {Tristan}}, \bibinfo {author} {\bibfnamefont {J.}~\bibnamefont {Hemberger}},
  \bibinfo {author} {\bibfnamefont {A.}~\bibnamefont {Krimmel}}, \bibinfo
  {author} {\bibfnamefont {H.~A.~K.}\ \bibnamefont {von Nidda}}, \bibinfo
  {author} {\bibfnamefont {V.}~\bibnamefont {Tsurkan}},\ and\ \bibinfo {author}
  {\bibfnamefont {A.}~\bibnamefont {Loidl}},\ }\bibfield  {title} {\bibinfo
  {title} {Geometric frustration in the cubic spinels mal2o4 (m=co, fe, and
  mn)},\ }\href {<Go to ISI>://WOS:000233603500070} {\bibfield  {journal}
  {\bibinfo  {journal} {Physical Review B}\ }\textbf {\bibinfo {volume} {72}}
  (\bibinfo {year} {2005})}\BibitemShut {NoStop}%
\bibitem [{\citenamefont {Suzuki}\ \emph {et~al.}(2007)\citenamefont {Suzuki},
  \citenamefont {Nagai}, \citenamefont {Nohara},\ and\ \citenamefont
  {Takagi}}]{Melting2007TS}%
  \BibitemOpen
  \bibfield  {author} {\bibinfo {author} {\bibfnamefont {T.}~\bibnamefont
  {Suzuki}}, \bibinfo {author} {\bibfnamefont {H.}~\bibnamefont {Nagai}},
  \bibinfo {author} {\bibfnamefont {M.}~\bibnamefont {Nohara}},\ and\ \bibinfo
  {author} {\bibfnamefont {H.}~\bibnamefont {Takagi}},\ }\bibfield  {title}
  {\bibinfo {title} {Melting of antiferromagnetic ordering in spinel oxide
  coal2o4},\ }\href {<Go to ISI>://WOS:000245670300066} {\bibfield  {journal}
  {\bibinfo  {journal} {Journal of Physics-Condensed Matter}\ }\textbf
  {\bibinfo {volume} {19}} (\bibinfo {year} {2007})}\BibitemShut {NoStop}%
\bibitem [{\citenamefont {Fritsch}\ \emph {et~al.}(2004)\citenamefont
  {Fritsch}, \citenamefont {Hemberger}, \citenamefont {Buttgen}, \citenamefont
  {Scheidt}, \citenamefont {von Nidda}, \citenamefont {Loidl},\ and\
  \citenamefont {Tsurkan}}]{Spin2004VH}%
  \BibitemOpen
  \bibfield  {author} {\bibinfo {author} {\bibfnamefont {V.}~\bibnamefont
  {Fritsch}}, \bibinfo {author} {\bibfnamefont {J.}~\bibnamefont {Hemberger}},
  \bibinfo {author} {\bibfnamefont {N.}~\bibnamefont {Buttgen}}, \bibinfo
  {author} {\bibfnamefont {E.~W.}\ \bibnamefont {Scheidt}}, \bibinfo {author}
  {\bibfnamefont {H.~A.~K.}\ \bibnamefont {von Nidda}}, \bibinfo {author}
  {\bibfnamefont {A.}~\bibnamefont {Loidl}},\ and\ \bibinfo {author}
  {\bibfnamefont {V.}~\bibnamefont {Tsurkan}},\ }\bibfield  {title} {\bibinfo
  {title} {Spin and orbital frustration in mnsc2s4 and fesc2s4},\ }\href {<Go
  to ISI>://WOS:000220344400039} {\bibfield  {journal} {\bibinfo  {journal}
  {Physical Review Letters}\ }\textbf {\bibinfo {volume} {92}} (\bibinfo {year}
  {2004})}\BibitemShut {NoStop}%
\bibitem [{\citenamefont {Chamorro}\ \emph {et~al.}(2018)\citenamefont
  {Chamorro}, \citenamefont {Ge}, \citenamefont {Flynn}, \citenamefont
  {Subramanian}, \citenamefont {Mourigal},\ and\ \citenamefont
  {McQueen}}]{Frustrated2018JC}%
  \BibitemOpen
  \bibfield  {author} {\bibinfo {author} {\bibfnamefont {J.~R.}\ \bibnamefont
  {Chamorro}}, \bibinfo {author} {\bibfnamefont {L.}~\bibnamefont {Ge}},
  \bibinfo {author} {\bibfnamefont {J.}~\bibnamefont {Flynn}}, \bibinfo
  {author} {\bibfnamefont {M.~A.}\ \bibnamefont {Subramanian}}, \bibinfo
  {author} {\bibfnamefont {M.}~\bibnamefont {Mourigal}},\ and\ \bibinfo
  {author} {\bibfnamefont {T.~M.}\ \bibnamefont {McQueen}},\ }\bibfield
  {title} {\bibinfo {title} {Frustrated spin one on a diamond lattice in
  nirh2o4},\ }\href {<Go to ISI>://WOS:000428509900001} {\bibfield  {journal}
  {\bibinfo  {journal} {Physical Review Materials}\ }\textbf {\bibinfo {volume}
  {2}} (\bibinfo {year} {2018})}\BibitemShut {NoStop}%
\bibitem [{\citenamefont {Liu}\ \emph {et~al.}(2019{\natexlab{a}})\citenamefont
  {Liu}, \citenamefont {Singh}, \citenamefont {Kirby}, \citenamefont {Zhong},
  \citenamefont {Cao}, \citenamefont {Pal}, \citenamefont {Kareev},
  \citenamefont {Middey}, \citenamefont {Freeland}, \citenamefont {Shafer},
  \citenamefont {Arenholz}, \citenamefont {Vanderbilt},\ and\ \citenamefont
  {Chakhalian}}]{Emergent2019XL}%
  \BibitemOpen
  \bibfield  {author} {\bibinfo {author} {\bibfnamefont {X.~R.}\ \bibnamefont
  {Liu}}, \bibinfo {author} {\bibfnamefont {S.}~\bibnamefont {Singh}}, \bibinfo
  {author} {\bibfnamefont {B.~J.}\ \bibnamefont {Kirby}}, \bibinfo {author}
  {\bibfnamefont {Z.~C.}\ \bibnamefont {Zhong}}, \bibinfo {author}
  {\bibfnamefont {Y.~W.}\ \bibnamefont {Cao}}, \bibinfo {author} {\bibfnamefont
  {B.}~\bibnamefont {Pal}}, \bibinfo {author} {\bibfnamefont {M.}~\bibnamefont
  {Kareev}}, \bibinfo {author} {\bibfnamefont {S.}~\bibnamefont {Middey}},
  \bibinfo {author} {\bibfnamefont {J.~W.}\ \bibnamefont {Freeland}}, \bibinfo
  {author} {\bibfnamefont {P.}~\bibnamefont {Shafer}}, \bibinfo {author}
  {\bibfnamefont {E.}~\bibnamefont {Arenholz}}, \bibinfo {author}
  {\bibfnamefont {D.}~\bibnamefont {Vanderbilt}},\ and\ \bibinfo {author}
  {\bibfnamefont {J.}~\bibnamefont {Chakhalian}},\ }\bibfield  {title}
  {\bibinfo {title} {Emergent magnetic state in (111)-oriented
  quasi-two-dimensional spinel oxides},\ }\href {<Go to
  ISI>://WOS:000502687500005} {\bibfield  {journal} {\bibinfo  {journal} {Nano
  Letters}\ }\textbf {\bibinfo {volume} {19}},\ \bibinfo {pages} {8381}
  (\bibinfo {year} {2019}{\natexlab{a}})}\BibitemShut {NoStop}%
\bibitem [{\citenamefont {Liu}\ \emph {et~al.}(2019{\natexlab{b}})\citenamefont
  {Liu}, \citenamefont {Asaba}, \citenamefont {Zhang}, \citenamefont {Cao},
  \citenamefont {Pal}, \citenamefont {Middey}, \citenamefont {Kumar},
  \citenamefont {Kareev}, \citenamefont {Gu}, \citenamefont {Sarma},
  \citenamefont {Shafer}, \citenamefont {Arenholz}, \citenamefont {Freeland},
  \citenamefont {Li},\ and\ \citenamefont {Chakhalian}}]{Quantum2020XL}%
  \BibitemOpen
  \bibfield  {author} {\bibinfo {author} {\bibfnamefont {X.}~\bibnamefont
  {Liu}}, \bibinfo {author} {\bibfnamefont {T.}~\bibnamefont {Asaba}}, \bibinfo
  {author} {\bibfnamefont {Q.}~\bibnamefont {Zhang}}, \bibinfo {author}
  {\bibfnamefont {Y.}~\bibnamefont {Cao}}, \bibinfo {author} {\bibfnamefont
  {B.}~\bibnamefont {Pal}}, \bibinfo {author} {\bibfnamefont {S.}~\bibnamefont
  {Middey}}, \bibinfo {author} {\bibfnamefont {P.}~\bibnamefont {Kumar}},
  \bibinfo {author} {\bibfnamefont {M.}~\bibnamefont {Kareev}}, \bibinfo
  {author} {\bibfnamefont {L.}~\bibnamefont {Gu}}, \bibinfo {author}
  {\bibfnamefont {D.}~\bibnamefont {Sarma}}, \bibinfo {author} {\bibfnamefont
  {P.}~\bibnamefont {Shafer}}, \bibinfo {author} {\bibfnamefont
  {E.}~\bibnamefont {Arenholz}}, \bibinfo {author} {\bibfnamefont
  {J.}~\bibnamefont {Freeland}}, \bibinfo {author} {\bibfnamefont
  {L.}~\bibnamefont {Li}},\ and\ \bibinfo {author} {\bibfnamefont
  {J.}~\bibnamefont {Chakhalian}},\ }\bibfield  {title} {\bibinfo {title}
  {Quantum spin liquids by geometric lattice design},\ }\href@noop {}
  {\bibfield  {journal} {\bibinfo  {journal} {arXiv preprint arXiv:1911.00100}\
  } (\bibinfo {year} {2019}{\natexlab{b}})}\BibitemShut {NoStop}%
\bibitem [{\citenamefont {Dutton}\ \emph {et~al.}(2011)\citenamefont {Dutton},
  \citenamefont {Huang}, \citenamefont {Tchernyshyov}, \citenamefont
  {Broholm},\ and\ \citenamefont {Cava}}]{Sensitivity2011SD}%
  \BibitemOpen
  \bibfield  {author} {\bibinfo {author} {\bibfnamefont {S.~E.}\ \bibnamefont
  {Dutton}}, \bibinfo {author} {\bibfnamefont {Q.}~\bibnamefont {Huang}},
  \bibinfo {author} {\bibfnamefont {O.}~\bibnamefont {Tchernyshyov}}, \bibinfo
  {author} {\bibfnamefont {C.~L.}\ \bibnamefont {Broholm}},\ and\ \bibinfo
  {author} {\bibfnamefont {R.~J.}\ \bibnamefont {Cava}},\ }\bibfield  {title}
  {\bibinfo {title} {Sensitivity of the magnetic properties of the zncr2o4 and
  mgcr2o4 spinels to nonstoichiometry},\ }\href {<Go to
  ISI>://WOS:000287364700006} {\bibfield  {journal} {\bibinfo  {journal}
  {Physical Review B}\ }\textbf {\bibinfo {volume} {83}} (\bibinfo {year}
  {2011})}\BibitemShut {NoStop}%
\bibitem [{\citenamefont {Ortega-San-Martin}\ \emph {et~al.}(2008)\citenamefont
  {Ortega-San-Martin}, \citenamefont {Williams}, \citenamefont {Gordon},
  \citenamefont {Klemme},\ and\ \citenamefont {Attfield}}]{Low2008LO}%
  \BibitemOpen
  \bibfield  {author} {\bibinfo {author} {\bibfnamefont {L.}~\bibnamefont
  {Ortega-San-Martin}}, \bibinfo {author} {\bibfnamefont {A.~J.}\ \bibnamefont
  {Williams}}, \bibinfo {author} {\bibfnamefont {C.~D.}\ \bibnamefont
  {Gordon}}, \bibinfo {author} {\bibfnamefont {S.}~\bibnamefont {Klemme}},\
  and\ \bibinfo {author} {\bibfnamefont {J.~P.}\ \bibnamefont {Attfield}},\
  }\bibfield  {title} {\bibinfo {title} {Low temperature neutron diffraction
  study of mgcr2o4 spinel},\ }\href {<Go to ISI>://WOS:000254101000039}
  {\bibfield  {journal} {\bibinfo  {journal} {Journal of Physics-Condensed
  Matter}\ }\textbf {\bibinfo {volume} {20}} (\bibinfo {year}
  {2008})}\BibitemShut {NoStop}%
\bibitem [{\citenamefont {McComb}\ \emph {et~al.}(2003)\citenamefont {McComb},
  \citenamefont {Craven}, \citenamefont {Chioncel}, \citenamefont
  {Lichtenstein},\ and\ \citenamefont {Docherty}}]{effect2003DM}%
  \BibitemOpen
  \bibfield  {author} {\bibinfo {author} {\bibfnamefont {D.~W.}\ \bibnamefont
  {McComb}}, \bibinfo {author} {\bibfnamefont {A.~J.}\ \bibnamefont {Craven}},
  \bibinfo {author} {\bibfnamefont {L.}~\bibnamefont {Chioncel}}, \bibinfo
  {author} {\bibfnamefont {A.~I.}\ \bibnamefont {Lichtenstein}},\ and\ \bibinfo
  {author} {\bibfnamefont {F.~T.}\ \bibnamefont {Docherty}},\ }\bibfield
  {title} {\bibinfo {title} {Effect of short-range magnetic ordering on
  electron energy-loss spectra in spinels},\ }\href {<Go to
  ISI>://WOS:000188081900065} {\bibfield  {journal} {\bibinfo  {journal}
  {Physical Review B}\ }\textbf {\bibinfo {volume} {68}} (\bibinfo {year}
  {2003})}\BibitemShut {NoStop}%
\bibitem [{\citenamefont {Rovers}\ \emph {et~al.}(2002)\citenamefont {Rovers},
  \citenamefont {Kyriakou}, \citenamefont {Dabkowska}, \citenamefont {Luke},
  \citenamefont {Larkin},\ and\ \citenamefont {Savici}}]{MT2002Muon}%
  \BibitemOpen
  \bibfield  {author} {\bibinfo {author} {\bibfnamefont {M.~T.}\ \bibnamefont
  {Rovers}}, \bibinfo {author} {\bibfnamefont {P.~P.}\ \bibnamefont
  {Kyriakou}}, \bibinfo {author} {\bibfnamefont {H.~A.}\ \bibnamefont
  {Dabkowska}}, \bibinfo {author} {\bibfnamefont {G.~M.}\ \bibnamefont {Luke}},
  \bibinfo {author} {\bibfnamefont {M.~I.}\ \bibnamefont {Larkin}},\ and\
  \bibinfo {author} {\bibfnamefont {A.~T.}\ \bibnamefont {Savici}},\ }\bibfield
   {title} {\bibinfo {title} {Muon-spin-relaxation investigation of the spin
  dynamics of geometrically frustrated chromium spinels},\ }\href {<Go to
  ISI>://WOS:000179611700078} {\bibfield  {journal} {\bibinfo  {journal}
  {Physical Review B}\ }\textbf {\bibinfo {volume} {66}} (\bibinfo {year}
  {2002})}\BibitemShut {NoStop}%
\bibitem [{\citenamefont {Suzuki}\ and\ \citenamefont
  {Tsunoda}(2007)}]{Spinel007HS}%
  \BibitemOpen
  \bibfield  {author} {\bibinfo {author} {\bibfnamefont {H.}~\bibnamefont
  {Suzuki}}\ and\ \bibinfo {author} {\bibfnamefont {Y.}~\bibnamefont
  {Tsunoda}},\ }\bibfield  {title} {\bibinfo {title} {Spinel-type frustrated
  system mgcr2o4 studied by neutron scattering and magnetization
  measurements},\ }\href {<Go to ISI>://WOS:000251594200013} {\bibfield
  {journal} {\bibinfo  {journal} {Journal of Physics and Chemistry of Solids}\
  }\textbf {\bibinfo {volume} {68}},\ \bibinfo {pages} {2060} (\bibinfo {year}
  {2007})}\BibitemShut {NoStop}%
\bibitem [{\citenamefont {Kemei}\ \emph {et~al.}(2013)\citenamefont {Kemei},
  \citenamefont {Barton}, \citenamefont {Moffitt}, \citenamefont {Gaultois},
  \citenamefont {Kurzman}, \citenamefont {Seshadri}, \citenamefont {Suchomel},\
  and\ \citenamefont {Kim}}]{crystal2013MK}%
  \BibitemOpen
  \bibfield  {author} {\bibinfo {author} {\bibfnamefont {M.~C.}\ \bibnamefont
  {Kemei}}, \bibinfo {author} {\bibfnamefont {P.~T.}\ \bibnamefont {Barton}},
  \bibinfo {author} {\bibfnamefont {S.~L.}\ \bibnamefont {Moffitt}}, \bibinfo
  {author} {\bibfnamefont {M.~W.}\ \bibnamefont {Gaultois}}, \bibinfo {author}
  {\bibfnamefont {J.~A.}\ \bibnamefont {Kurzman}}, \bibinfo {author}
  {\bibfnamefont {R.}~\bibnamefont {Seshadri}}, \bibinfo {author}
  {\bibfnamefont {M.~R.}\ \bibnamefont {Suchomel}},\ and\ \bibinfo {author}
  {\bibfnamefont {Y.~I.}\ \bibnamefont {Kim}},\ }\bibfield  {title} {\bibinfo
  {title} {Crystal structures of spin-jahn-teller-ordered mgcr2o4 and
  zncr2o4},\ }\href {<Go to ISI>://WOS:000322227600013} {\bibfield  {journal}
  {\bibinfo  {journal} {Journal of Physics-Condensed Matter}\ }\textbf
  {\bibinfo {volume} {25}} (\bibinfo {year} {2013})}\BibitemShut {NoStop}%
\bibitem [{\citenamefont {Tomiyasu}\ \emph {et~al.}(2008)\citenamefont
  {Tomiyasu}, \citenamefont {Suzuki}, \citenamefont {Toki}, \citenamefont
  {Itoh}, \citenamefont {Matsuura}, \citenamefont {Aso},\ and\ \citenamefont
  {Yamada}}]{Molecular2008KT}%
  \BibitemOpen
  \bibfield  {author} {\bibinfo {author} {\bibfnamefont {K.}~\bibnamefont
  {Tomiyasu}}, \bibinfo {author} {\bibfnamefont {H.}~\bibnamefont {Suzuki}},
  \bibinfo {author} {\bibfnamefont {M.}~\bibnamefont {Toki}}, \bibinfo {author}
  {\bibfnamefont {S.}~\bibnamefont {Itoh}}, \bibinfo {author} {\bibfnamefont
  {M.}~\bibnamefont {Matsuura}}, \bibinfo {author} {\bibfnamefont
  {N.}~\bibnamefont {Aso}},\ and\ \bibinfo {author} {\bibfnamefont
  {K.}~\bibnamefont {Yamada}},\ }\bibfield  {title} {\bibinfo {title}
  {Molecular spin resonance in the geometrically frustrated magnet mgcr2o4 by
  inelastic neutron scattering},\ }\href@noop {} {\bibfield  {journal}
  {\bibinfo  {journal} {Physical review letters}\ }\textbf {\bibinfo {volume}
  {101}},\ \bibinfo {pages} {177401} (\bibinfo {year} {2008})}\BibitemShut
  {NoStop}%
\bibitem [{\citenamefont {Tomiyasu}\ \emph {et~al.}(2013)\citenamefont
  {Tomiyasu}, \citenamefont {Yokobori}, \citenamefont {Kousaka}, \citenamefont
  {Bewley}, \citenamefont {Guidi}, \citenamefont {Watanabe}, \citenamefont
  {Akimitsu},\ and\ \citenamefont {Yamada}}]{Emergence2013KT}%
  \BibitemOpen
  \bibfield  {author} {\bibinfo {author} {\bibfnamefont {K.}~\bibnamefont
  {Tomiyasu}}, \bibinfo {author} {\bibfnamefont {T.}~\bibnamefont {Yokobori}},
  \bibinfo {author} {\bibfnamefont {Y.}~\bibnamefont {Kousaka}}, \bibinfo
  {author} {\bibfnamefont {R.~I.}\ \bibnamefont {Bewley}}, \bibinfo {author}
  {\bibfnamefont {T.}~\bibnamefont {Guidi}}, \bibinfo {author} {\bibfnamefont
  {T.}~\bibnamefont {Watanabe}}, \bibinfo {author} {\bibfnamefont
  {J.}~\bibnamefont {Akimitsu}},\ and\ \bibinfo {author} {\bibfnamefont
  {K.}~\bibnamefont {Yamada}},\ }\bibfield  {title} {\bibinfo {title}
  {Emergence of highly degenerate excited states in the frustrated magnet
  mgcr2o4},\ }\href {<Go to ISI>://WOS:000314995400009} {\bibfield  {journal}
  {\bibinfo  {journal} {Physical Review Letters}\ }\textbf {\bibinfo {volume}
  {110}} (\bibinfo {year} {2013})}\BibitemShut {NoStop}%
\bibitem [{\citenamefont {Miyata}\ \emph {et~al.}(2014)\citenamefont {Miyata},
  \citenamefont {Ueda},\ and\ \citenamefont {Takeyama}}]{Canted2014AM}%
  \BibitemOpen
  \bibfield  {author} {\bibinfo {author} {\bibfnamefont {A.}~\bibnamefont
  {Miyata}}, \bibinfo {author} {\bibfnamefont {H.}~\bibnamefont {Ueda}},\ and\
  \bibinfo {author} {\bibfnamefont {S.}~\bibnamefont {Takeyama}},\ }\bibfield
  {title} {\bibinfo {title} {Canted 2: 1: 1 magnetic supersolid phase in a
  frustrated magnet mgcr2o4 as a small limit of the biquadratic spin
  interaction},\ }\href@noop {} {\bibfield  {journal} {\bibinfo  {journal}
  {Journal of the Physical Society of Japan}\ }\textbf {\bibinfo {volume}
  {83}},\ \bibinfo {pages} {063702} (\bibinfo {year} {2014})}\BibitemShut
  {NoStop}%
\bibitem [{\citenamefont {Bai}\ \emph {et~al.}(2019)\citenamefont {Bai},
  \citenamefont {Paddison}, \citenamefont {Kapit}, \citenamefont {Koohpayeh},
  \citenamefont {Wen}, \citenamefont {Dutton}, \citenamefont {Savici},
  \citenamefont {Kolesnikov}, \citenamefont {Granroth}, \citenamefont
  {Broholm}, \citenamefont {Chalker},\ and\ \citenamefont
  {Mourigal}}]{magnetic2019SB}%
  \BibitemOpen
  \bibfield  {author} {\bibinfo {author} {\bibfnamefont {X.}~\bibnamefont
  {Bai}}, \bibinfo {author} {\bibfnamefont {J.~A.~M.}\ \bibnamefont
  {Paddison}}, \bibinfo {author} {\bibfnamefont {E.}~\bibnamefont {Kapit}},
  \bibinfo {author} {\bibfnamefont {S.~M.}\ \bibnamefont {Koohpayeh}}, \bibinfo
  {author} {\bibfnamefont {J.~J.}\ \bibnamefont {Wen}}, \bibinfo {author}
  {\bibfnamefont {S.~E.}\ \bibnamefont {Dutton}}, \bibinfo {author}
  {\bibfnamefont {A.~T.}\ \bibnamefont {Savici}}, \bibinfo {author}
  {\bibfnamefont {A.~I.}\ \bibnamefont {Kolesnikov}}, \bibinfo {author}
  {\bibfnamefont {G.~E.}\ \bibnamefont {Granroth}}, \bibinfo {author}
  {\bibfnamefont {C.~L.}\ \bibnamefont {Broholm}}, \bibinfo {author}
  {\bibfnamefont {J.~T.}\ \bibnamefont {Chalker}},\ and\ \bibinfo {author}
  {\bibfnamefont {M.}~\bibnamefont {Mourigal}},\ }\bibfield  {title} {\bibinfo
  {title} {Magnetic excitations of the classical spin liquid mgcr2o4},\ }\href
  {<Go to ISI>://WOS:000460669200008} {\bibfield  {journal} {\bibinfo
  {journal} {Physical Review Letters}\ }\textbf {\bibinfo {volume} {122}}
  (\bibinfo {year} {2019})}\BibitemShut {NoStop}%
\bibitem [{\citenamefont {Luders}\ \emph {et~al.}(2004)\citenamefont {Luders},
  \citenamefont {Sanchez},\ and\ \citenamefont {Fontcuberta}}]{Initial2004UL}%
  \BibitemOpen
  \bibfield  {author} {\bibinfo {author} {\bibfnamefont {U.}~\bibnamefont
  {Luders}}, \bibinfo {author} {\bibfnamefont {F.}~\bibnamefont {Sanchez}},\
  and\ \bibinfo {author} {\bibfnamefont {J.}~\bibnamefont {Fontcuberta}},\
  }\bibfield  {title} {\bibinfo {title} {Initial stages in the growth of
  {111}-faceted cocr2o4 clusters: mechanisms and strained nanometric
  pyramids},\ }\href {<Go to ISI>://WOS:000220385600016} {\bibfield  {journal}
  {\bibinfo  {journal} {Applied Physics a-Materials Science \& Processing}\
  }\textbf {\bibinfo {volume} {79}},\ \bibinfo {pages} {93} (\bibinfo {year}
  {2004})}\BibitemShut {NoStop}%
\bibitem [{\citenamefont {Chakhalian}\ \emph {et~al.}(2020)\citenamefont
  {Chakhalian}, \citenamefont {Liu},\ and\ \citenamefont
  {Fiete}}]{Strongly2020JC}%
  \BibitemOpen
  \bibfield  {author} {\bibinfo {author} {\bibfnamefont {J.}~\bibnamefont
  {Chakhalian}}, \bibinfo {author} {\bibfnamefont {X.~R.}\ \bibnamefont
  {Liu}},\ and\ \bibinfo {author} {\bibfnamefont {G.~A.}\ \bibnamefont
  {Fiete}},\ }\bibfield  {title} {\bibinfo {title} {Strongly correlated and
  topological states in 111 grown transition metal oxide thin films and
  heterostructures},\ }\href {<Go to ISI>://WOS:000535981100001} {\bibfield
  {journal} {\bibinfo  {journal} {Apl Materials}\ }\textbf {\bibinfo {volume}
  {8}} (\bibinfo {year} {2020})}\BibitemShut {NoStop}%
\bibitem [{\citenamefont {Liu}\ \emph {et~al.}(2016)\citenamefont {Liu},
  \citenamefont {Middey}, \citenamefont {Cao}, \citenamefont {Kareev},\ and\
  \citenamefont {Chakhalian}}]{Geometrical2016XL}%
  \BibitemOpen
  \bibfield  {author} {\bibinfo {author} {\bibfnamefont {X.~R.}\ \bibnamefont
  {Liu}}, \bibinfo {author} {\bibfnamefont {S.}~\bibnamefont {Middey}},
  \bibinfo {author} {\bibfnamefont {Y.~W.}\ \bibnamefont {Cao}}, \bibinfo
  {author} {\bibfnamefont {M.}~\bibnamefont {Kareev}},\ and\ \bibinfo {author}
  {\bibfnamefont {J.}~\bibnamefont {Chakhalian}},\ }\bibfield  {title}
  {\bibinfo {title} {Geometrical lattice engineering of complex oxide
  heterostructures: a designer approach to emergent quantum states},\ }\href
  {<Go to ISI>://WOS:000389137400002} {\bibfield  {journal} {\bibinfo
  {journal} {Mrs Communications}\ }\textbf {\bibinfo {volume} {6}},\ \bibinfo
  {pages} {133} (\bibinfo {year} {2016})}\BibitemShut {NoStop}%
\bibitem [{\citenamefont {Grimes}(1971)}]{Structural1971NWG}%
  \BibitemOpen
  \bibfield  {author} {\bibinfo {author} {\bibfnamefont {N.~W.}\ \bibnamefont
  {Grimes}},\ }\bibfield  {title} {\bibinfo {title} {Structural distortions in
  mgcr2o4},\ }\href {<Go to ISI>://WOS:A1971K980600006} {\bibfield  {journal}
  {\bibinfo  {journal} {Journal of Physics Part C Solid State Physics}\
  }\textbf {\bibinfo {volume} {4}},\ \bibinfo {pages} {L342} (\bibinfo {year}
  {1971})}\BibitemShut {NoStop}%
\bibitem [{\citenamefont {Kosenko}\ \emph {et~al.}(2020)\citenamefont
  {Kosenko}, \citenamefont {Filatova},\ and\ \citenamefont
  {Egorova}}]{magnesiochromate2020NFK}%
  \BibitemOpen
  \bibfield  {author} {\bibinfo {author} {\bibfnamefont {N.~F.}\ \bibnamefont
  {Kosenko}}, \bibinfo {author} {\bibfnamefont {N.~V.}\ \bibnamefont
  {Filatova}},\ and\ \bibinfo {author} {\bibfnamefont {A.~A.}\ \bibnamefont
  {Egorova}},\ }\bibfield  {title} {\bibinfo {title} {Magnesiochromite
  (mgcr$_2$o$_4$) synthesis: Effect of mechanical and microwave pretreatment},\
  }\href {<Go to ISI>://WOS:000546901000009} {\bibfield  {journal} {\bibinfo
  {journal} {Izvestiya Vysshikh Uchebnykh Zavedenii Khimiya I Khimicheskaya
  Tekhnologiya}\ }\textbf {\bibinfo {volume} {63}},\ \bibinfo {pages} {96}
  (\bibinfo {year} {2020})}\BibitemShut {NoStop}%
\bibitem [{\citenamefont {Hu}\ \emph {et~al.}(2014)\citenamefont {Hu},
  \citenamefont {Zhao}, \citenamefont {Hu}, \citenamefont {Chang},
  \citenamefont {Li},\ and\ \citenamefont {Wang}}]{Catalytic2014JH}%
  \BibitemOpen
  \bibfield  {author} {\bibinfo {author} {\bibfnamefont {J.~N.}\ \bibnamefont
  {Hu}}, \bibinfo {author} {\bibfnamefont {W.~Y.}\ \bibnamefont {Zhao}},
  \bibinfo {author} {\bibfnamefont {R.~S.}\ \bibnamefont {Hu}}, \bibinfo
  {author} {\bibfnamefont {G.~Y.}\ \bibnamefont {Chang}}, \bibinfo {author}
  {\bibfnamefont {C.}~\bibnamefont {Li}},\ and\ \bibinfo {author}
  {\bibfnamefont {L.~J.}\ \bibnamefont {Wang}},\ }\bibfield  {title} {\bibinfo
  {title} {Catalytic activity of spinel oxides mgcr2o4 and cocr2o4 for methane
  combustion},\ }\href {<Go to ISI>://WOS:000340313400043} {\bibfield
  {journal} {\bibinfo  {journal} {Materials Research Bulletin}\ }\textbf
  {\bibinfo {volume} {57}},\ \bibinfo {pages} {268} (\bibinfo {year}
  {2014})}\BibitemShut {NoStop}%
\bibitem [{\citenamefont {Jafarnejad}\ \emph {et~al.}(2016)\citenamefont
  {Jafarnejad}, \citenamefont {Khanahmadzadeh}, \citenamefont {Ghanbary},\ and\
  \citenamefont {Enhessari}}]{Synthesis2016EJ}%
  \BibitemOpen
  \bibfield  {author} {\bibinfo {author} {\bibfnamefont {E.}~\bibnamefont
  {Jafarnejad}}, \bibinfo {author} {\bibfnamefont {S.}~\bibnamefont
  {Khanahmadzadeh}}, \bibinfo {author} {\bibfnamefont {F.}~\bibnamefont
  {Ghanbary}},\ and\ \bibinfo {author} {\bibfnamefont {M.}~\bibnamefont
  {Enhessari}},\ }\bibfield  {title} {\bibinfo {title} {Synthesis,
  characterization and optical band gap of pirochromite (mgcr2o4) nanoparticles
  by stearic acid sol-gel method},\ }\href@noop {} {\bibfield  {journal}
  {\bibinfo  {journal} {Current Chemistry Letters}\ }\textbf {\bibinfo {volume}
  {5}},\ \bibinfo {pages} {173} (\bibinfo {year} {2016})}\BibitemShut {NoStop}%
\bibitem [{\citenamefont {Koohpayeh}\ \emph {et~al.}(2013)\citenamefont
  {Koohpayeh}, \citenamefont {Wen}, \citenamefont {Mourigal}, \citenamefont
  {Dutton}, \citenamefont {Cava}, \citenamefont {Broholm},\ and\ \citenamefont
  {McQueen}}]{optical2013SK}%
  \BibitemOpen
  \bibfield  {author} {\bibinfo {author} {\bibfnamefont {S.~M.}\ \bibnamefont
  {Koohpayeh}}, \bibinfo {author} {\bibfnamefont {J.~J.}\ \bibnamefont {Wen}},
  \bibinfo {author} {\bibfnamefont {M.}~\bibnamefont {Mourigal}}, \bibinfo
  {author} {\bibfnamefont {S.~E.}\ \bibnamefont {Dutton}}, \bibinfo {author}
  {\bibfnamefont {R.~J.}\ \bibnamefont {Cava}}, \bibinfo {author}
  {\bibfnamefont {C.~L.}\ \bibnamefont {Broholm}},\ and\ \bibinfo {author}
  {\bibfnamefont {T.~M.}\ \bibnamefont {McQueen}},\ }\bibfield  {title}
  {\bibinfo {title} {Optical floating zone crystal growth and magnetic
  properties of mgcr2o4},\ }\href {<Go to ISI>://WOS:000326680900008}
  {\bibfield  {journal} {\bibinfo  {journal} {Journal of Crystal Growth}\
  }\textbf {\bibinfo {volume} {384}},\ \bibinfo {pages} {39} (\bibinfo {year}
  {2013})}\BibitemShut {NoStop}%
\bibitem [{\citenamefont {Durrani}\ \emph {et~al.}(2014)\citenamefont
  {Durrani}, \citenamefont {Naz}, \citenamefont {Nadeem},\ and\ \citenamefont
  {Khan}}]{Thermal2014SK}%
  \BibitemOpen
  \bibfield  {author} {\bibinfo {author} {\bibfnamefont {S.~K.}\ \bibnamefont
  {Durrani}}, \bibinfo {author} {\bibfnamefont {S.}~\bibnamefont {Naz}},
  \bibinfo {author} {\bibfnamefont {M.}~\bibnamefont {Nadeem}},\ and\ \bibinfo
  {author} {\bibfnamefont {A.~A.}\ \bibnamefont {Khan}},\ }\bibfield  {title}
  {\bibinfo {title} {Thermal, structural, and impedance analysis of
  nanocrystalline magnesium chromite spinel synthesized via hydrothermal
  process},\ }\href {<Go to ISI>://WOS:000333533000037} {\bibfield  {journal}
  {\bibinfo  {journal} {Journal of Thermal Analysis and Calorimetry}\ }\textbf
  {\bibinfo {volume} {116}},\ \bibinfo {pages} {309} (\bibinfo {year}
  {2014})}\BibitemShut {NoStop}%
\bibitem [{\citenamefont {Morozova}\ and\ \citenamefont
  {Popov}(2010)}]{synthesis2010LM}%
  \BibitemOpen
  \bibfield  {author} {\bibinfo {author} {\bibfnamefont {L.~V.}\ \bibnamefont
  {Morozova}}\ and\ \bibinfo {author} {\bibfnamefont {V.~P.}\ \bibnamefont
  {Popov}},\ }\bibfield  {title} {\bibinfo {title} {Synthesis and investigation
  of magnesium chromium spinel},\ }\href {<Go to ISI>://WOS:000275461000015}
  {\bibfield  {journal} {\bibinfo  {journal} {Glass Physics and Chemistry}\
  }\textbf {\bibinfo {volume} {36}},\ \bibinfo {pages} {86} (\bibinfo {year}
  {2010})}\BibitemShut {NoStop}%
\bibitem [{\citenamefont {Bindi}\ \emph {et~al.}(2014)\citenamefont {Bindi},
  \citenamefont {Sirotkina}, \citenamefont {Bobrov},\ and\ \citenamefont
  {Irifune}}]{bindi2014x}%
  \BibitemOpen
  \bibfield  {author} {\bibinfo {author} {\bibfnamefont {L.}~\bibnamefont
  {Bindi}}, \bibinfo {author} {\bibfnamefont {E.}~\bibnamefont {Sirotkina}},
  \bibinfo {author} {\bibfnamefont {A.~V.}\ \bibnamefont {Bobrov}},\ and\
  \bibinfo {author} {\bibfnamefont {T.}~\bibnamefont {Irifune}},\ }\bibfield
  {title} {\bibinfo {title} {X-ray single-crystal structural characterization
  of mgcr2o4, a post-spinel phase synthesized at 23 gpa and 1600 c},\
  }\href@noop {} {\bibfield  {journal} {\bibinfo  {journal} {Journal of Physics
  and Chemistry of Solids}\ }\textbf {\bibinfo {volume} {75}},\ \bibinfo
  {pages} {638} (\bibinfo {year} {2014})}\BibitemShut {NoStop}%
\bibitem [{\citenamefont {Rasmussen}\ \emph {et~al.}(2012)\citenamefont
  {Rasmussen}, \citenamefont {Meinander}, \citenamefont {Besenbacher},\ and\
  \citenamefont {Lauritsen}}]{Noncontact2012BJN}%
  \BibitemOpen
  \bibfield  {author} {\bibinfo {author} {\bibfnamefont {M.~K.}\ \bibnamefont
  {Rasmussen}}, \bibinfo {author} {\bibfnamefont {K.}~\bibnamefont
  {Meinander}}, \bibinfo {author} {\bibfnamefont {F.}~\bibnamefont
  {Besenbacher}},\ and\ \bibinfo {author} {\bibfnamefont {J.~V.}\ \bibnamefont
  {Lauritsen}},\ }\bibfield  {title} {\bibinfo {title} {Noncontact atomic force
  microscopy study of the spinel mgal2o4(111) surface},\ }\href {<Go to
  ISI>://WOS:000301367600001} {\bibfield  {journal} {\bibinfo  {journal}
  {Beilstein Journal of Nanotechnology}\ }\textbf {\bibinfo {volume} {3}},\
  \bibinfo {pages} {192} (\bibinfo {year} {2012})}\BibitemShut {NoStop}%
\bibitem [{\citenamefont {Daneu}\ \emph {et~al.}(2007)\citenamefont {Daneu},
  \citenamefont {Recnik}, \citenamefont {Yamazaki},\ and\ \citenamefont
  {Dolenec}}]{structure2007ND}%
  \BibitemOpen
  \bibfield  {author} {\bibinfo {author} {\bibfnamefont {N.}~\bibnamefont
  {Daneu}}, \bibinfo {author} {\bibfnamefont {A.}~\bibnamefont {Recnik}},
  \bibinfo {author} {\bibfnamefont {T.}~\bibnamefont {Yamazaki}},\ and\
  \bibinfo {author} {\bibfnamefont {T.}~\bibnamefont {Dolenec}},\ }\bibfield
  {title} {\bibinfo {title} {Structure and chemistry of (111) twin boundaries
  in mgal2o4 spinel crystals from mogok},\ }\href {<Go to
  ISI>://WOS:000246099900003} {\bibfield  {journal} {\bibinfo  {journal}
  {Physics and Chemistry of Minerals}\ }\textbf {\bibinfo {volume} {34}},\
  \bibinfo {pages} {233} (\bibinfo {year} {2007})}\BibitemShut {NoStop}%
\bibitem [{\citenamefont {Noh}\ \emph {et~al.}(2014)\citenamefont {Noh},
  \citenamefont {Jeong}, \citenamefont {Chang}, \citenamefont {Jeong},
  \citenamefont {Moon}, \citenamefont {Cho}, \citenamefont {Ok}, \citenamefont
  {Kim}, \citenamefont {Kim}, \citenamefont {Min}, \citenamefont {Lee},
  \citenamefont {Kim}, \citenamefont {Park}, \citenamefont {Kim},\ and\
  \citenamefont {Lee}}]{Direct2014HN}%
  \BibitemOpen
  \bibfield  {author} {\bibinfo {author} {\bibfnamefont {H.~J.}\ \bibnamefont
  {Noh}}, \bibinfo {author} {\bibfnamefont {J.}~\bibnamefont {Jeong}}, \bibinfo
  {author} {\bibfnamefont {B.}~\bibnamefont {Chang}}, \bibinfo {author}
  {\bibfnamefont {D.}~\bibnamefont {Jeong}}, \bibinfo {author} {\bibfnamefont
  {H.~S.}\ \bibnamefont {Moon}}, \bibinfo {author} {\bibfnamefont {E.~J.}\
  \bibnamefont {Cho}}, \bibinfo {author} {\bibfnamefont {J.~M.}\ \bibnamefont
  {Ok}}, \bibinfo {author} {\bibfnamefont {J.~S.}\ \bibnamefont {Kim}},
  \bibinfo {author} {\bibfnamefont {K.}~\bibnamefont {Kim}}, \bibinfo {author}
  {\bibfnamefont {B.~I.}\ \bibnamefont {Min}}, \bibinfo {author} {\bibfnamefont
  {H.~K.}\ \bibnamefont {Lee}}, \bibinfo {author} {\bibfnamefont {J.~Y.}\
  \bibnamefont {Kim}}, \bibinfo {author} {\bibfnamefont {B.~G.}\ \bibnamefont
  {Park}}, \bibinfo {author} {\bibfnamefont {H.~D.}\ \bibnamefont {Kim}},\ and\
  \bibinfo {author} {\bibfnamefont {S.}~\bibnamefont {Lee}},\ }\bibfield
  {title} {\bibinfo {title} {Direct observation of localized spin
  antiferromagnetic transition in pdcro2 by angle-resolved photoemission
  spectroscopy},\ }\href {<Go to ISI>://WOS:000329846100009} {\bibfield
  {journal} {\bibinfo  {journal} {Scientific Reports}\ }\textbf {\bibinfo
  {volume} {4}} (\bibinfo {year} {2014})}\BibitemShut {NoStop}%
\bibitem [{\citenamefont {Biesinger}\ \emph {et~al.}(2004)\citenamefont
  {Biesinger}, \citenamefont {Brown}, \citenamefont {Mycroft}, \citenamefont
  {Davidson},\ and\ \citenamefont {McIntyre}}]{X-ray2004MB}%
  \BibitemOpen
  \bibfield  {author} {\bibinfo {author} {\bibfnamefont {M.~C.}\ \bibnamefont
  {Biesinger}}, \bibinfo {author} {\bibfnamefont {C.}~\bibnamefont {Brown}},
  \bibinfo {author} {\bibfnamefont {J.~R.}\ \bibnamefont {Mycroft}}, \bibinfo
  {author} {\bibfnamefont {R.~D.}\ \bibnamefont {Davidson}},\ and\ \bibinfo
  {author} {\bibfnamefont {N.~S.}\ \bibnamefont {McIntyre}},\ }\bibfield
  {title} {\bibinfo {title} {X-ray photoelectron spectroscopy studies of
  chromium compounds},\ }\href {<Go to ISI>://WOS:000225911200007} {\bibfield
  {journal} {\bibinfo  {journal} {Surface and Interface Analysis}\ }\textbf
  {\bibinfo {volume} {36}},\ \bibinfo {pages} {1550} (\bibinfo {year}
  {2004})}\BibitemShut {NoStop}%
\bibitem [{\citenamefont {Payne}\ \emph {et~al.}(2011)\citenamefont {Payne},
  \citenamefont {Biesinger},\ and\ \citenamefont {McIntyre}}]{X-ray2011BP}%
  \BibitemOpen
  \bibfield  {author} {\bibinfo {author} {\bibfnamefont {B.~P.}\ \bibnamefont
  {Payne}}, \bibinfo {author} {\bibfnamefont {M.~C.}\ \bibnamefont
  {Biesinger}},\ and\ \bibinfo {author} {\bibfnamefont {N.~S.}\ \bibnamefont
  {McIntyre}},\ }\bibfield  {title} {\bibinfo {title} {X-ray photoelectron
  spectroscopy studies of reactions on chromium metal and chromium oxide
  surfaces},\ }\href {<Go to ISI>://WOS:000289015400006} {\bibfield  {journal}
  {\bibinfo  {journal} {Journal of Electron Spectroscopy and Related
  Phenomena}\ }\textbf {\bibinfo {volume} {184}},\ \bibinfo {pages} {29}
  (\bibinfo {year} {2011})}\BibitemShut {NoStop}%
\bibitem [{\citenamefont {Kaul}\ \emph {et~al.}(2004)\citenamefont {Kaul},
  \citenamefont {Gorbenko},\ and\ \citenamefont {Kamenev}}]{Therole2004AK}%
  \BibitemOpen
  \bibfield  {author} {\bibinfo {author} {\bibfnamefont {A.~R.}\ \bibnamefont
  {Kaul}}, \bibinfo {author} {\bibfnamefont {O.~Y.}\ \bibnamefont {Gorbenko}},\
  and\ \bibinfo {author} {\bibfnamefont {A.~A.}\ \bibnamefont {Kamenev}},\
  }\bibfield  {title} {\bibinfo {title} {The role of heteroepitaxy in the
  development of new thin-film oxide-based functional materials},\ }\href {<Go
  to ISI>://WOS:000224563100003} {\bibfield  {journal} {\bibinfo  {journal}
  {Uspekhi Khimii}\ }\textbf {\bibinfo {volume} {73}},\ \bibinfo {pages} {932}
  (\bibinfo {year} {2004})}\BibitemShut {NoStop}%
\bibitem [{\citenamefont {Jacob}(1977)}]{Potentiometric1977KT}%
  \BibitemOpen
  \bibfield  {author} {\bibinfo {author} {\bibfnamefont {K.~T.}\ \bibnamefont
  {Jacob}},\ }\bibfield  {title} {\bibinfo {title} {Potentiometric
  determination of the gibbs free energy of formation of cadmium and magnesium
  chromites},\ }\href {<Go to ISI>://WOS:000208067600005} {\bibfield  {journal}
  {\bibinfo  {journal} {Journal of the Electrochemical Society}\ }\textbf
  {\bibinfo {volume} {124}},\ \bibinfo {pages} {1827} (\bibinfo {year}
  {1977})}\BibitemShut {NoStop}%
\bibitem [{\citenamefont {Mishra}\ and\ \citenamefont
  {Thomas}(1977)}]{Surface1977RM}%
  \BibitemOpen
  \bibfield  {author} {\bibinfo {author} {\bibfnamefont {R.~K.}\ \bibnamefont
  {Mishra}}\ and\ \bibinfo {author} {\bibfnamefont {G.}~\bibnamefont
  {Thomas}},\ }\bibfield  {title} {\bibinfo {title} {Surface-energy of
  spinel},\ }\href {<Go to ISI>://WOS:A1977EB23600027} {\bibfield  {journal}
  {\bibinfo  {journal} {Journal of Applied Physics}\ }\textbf {\bibinfo
  {volume} {48}},\ \bibinfo {pages} {4576} (\bibinfo {year}
  {1977})}\BibitemShut {NoStop}%
\bibitem [{\citenamefont {Watanabe}\ \emph {et~al.}(2012)\citenamefont
  {Watanabe}, \citenamefont {Ishikawa}, \citenamefont {Suzuki}, \citenamefont
  {Kousaka},\ and\ \citenamefont {Tomiyasu}}]{observation2012TW}%
  \BibitemOpen
  \bibfield  {author} {\bibinfo {author} {\bibfnamefont {T.}~\bibnamefont
  {Watanabe}}, \bibinfo {author} {\bibfnamefont {S.}~\bibnamefont {Ishikawa}},
  \bibinfo {author} {\bibfnamefont {H.}~\bibnamefont {Suzuki}}, \bibinfo
  {author} {\bibfnamefont {Y.}~\bibnamefont {Kousaka}},\ and\ \bibinfo {author}
  {\bibfnamefont {K.}~\bibnamefont {Tomiyasu}},\ }\bibfield  {title} {\bibinfo
  {title} {Observation of elastic anomalies driven by coexisting dynamical spin
  jahn-teller effect and dynamical molecular-spin state in the paramagnetic
  phase of frustrated mgcr2o4},\ }\href {<Go to ISI>://WOS:000310129600006}
  {\bibfield  {journal} {\bibinfo  {journal} {Physical Review B}\ }\textbf
  {\bibinfo {volume} {86}} (\bibinfo {year} {2012})}\BibitemShut {NoStop}%
\bibitem [{\citenamefont {Lal}\ and\ \citenamefont
  {Pandey}(2016)}]{therole2016SL}%
  \BibitemOpen
  \bibfield  {author} {\bibinfo {author} {\bibfnamefont {S.}~\bibnamefont
  {Lal}}\ and\ \bibinfo {author} {\bibfnamefont {S.~K.}\ \bibnamefont
  {Pandey}},\ }\bibfield  {title} {\bibinfo {title} {The role of ionic sizes in
  inducing the cubic to tetragonal distortion in av2o4 and acr2o4 (a= zn, mg
  and cd) compounds},\ }\href@noop {} {\bibfield  {journal} {\bibinfo
  {journal} {Materials Research Express}\ }\textbf {\bibinfo {volume} {3}},\
  \bibinfo {pages} {116301} (\bibinfo {year} {2016})}\BibitemShut {NoStop}%
\end{thebibliography}%

\date{\today}

\end{document}